\PassOptionsToPackage{dvipdfm}{graphicx}
\documentclass{comjnl}

\newcommand{\tbl}[1]
{
  \caption{{#1}}
} 

\usepackage[cmex10]{amsmath}

\usepackage{array}
\newcolumntype{C}[1]{>{\centering\arraybackslash}m{#1}}
\newcolumntype{L}[1]{>{\raggedright\arraybackslash}m{#1}}
\newcolumntype{R}[1]{>{\raggedleft\arraybackslash}m{#1}}

\usepackage{mdwmath}
\usepackage{eqparbox}
\usepackage[tight,footnotesize]{subfigure}
\usepackage{fixltx2e}
\usepackage{dblfloatfix}
\usepackage{lscape,longtable}
\usepackage{url}
\usepackage{balance}
\pdfpageattr {/Group << /S /Transparency /I true /CS /DeviceRGB>>}   
\usepackage{multicol}
\usepackage[x11names,table]{xcolor}
\usepackage{color, colortbl}
\usepackage{floatrow}
\newfloatcommand{capbtabbox}{table}[][\FBwidth]
\usepackage{blindtext}
\usepackage{textcomp}
\usepackage{graphicx}
\usepackage{lmodern}
\usepackage{slashbox}

\newcolumntype{L}{>{\centering\arraybackslash}m{3cm}}

\definecolor{Maroon}{rgb}{0.8, 0.0, 0.0}
\definecolor{gray}{rgb}{0.3, 0.3, 0.3}

\definecolor{rrrreducm}{cmyk}{0,1,0.71,0.21}
\definecolor{rrrblue}{cmyk}{0.86,0.33,0,0.50}
\definecolor{rrrgreen}{cmyk}{0.86,0.3,0.86,0.5}
\definecolor{rrrsilver}{cmyk}{0,0,0,0.1}
\definecolor{rrrsilver2}{cmyk}{0,0,0,0.07}
\definecolor{rrrred}{cmyk}{0,0.88,0.94,0.38}
\definecolor{rrrred2}{cmyk}{0,0.80,0.81,0.25}
\definecolor{rrrgreen2}{cmyk}{0.35,0,0.67,0.41}
\definecolor{rrrgreen3}{cmyk}{0.43,0,0.60,0.60}
\definecolor{rrrgray}{cmyk}{0,0,0,0.50}
\definecolor{rrrorange}{cmyk}{0,0.62,0.99,0}
\definecolor{rrryellow}{cmyk}{0,0.09,0.62,0.11}
\definecolor{rrrturquoise}{cmyk}{0.93,0,0.21,0.13}
\definecolor{rrrpurple}{cmyk}{0.33,0.87,0,0.06}
\definecolor{rrrbrown}{cmyk}{0,0.67,0.67,0.50}
\definecolor{rrrinvisible}{cmyk}{0,0,0,0}
\definecolor{rrrsgilightblue}{cmyk}{0.88,0.44,0,0.45}

\usepackage{bibentry}
\usepackage[english]{babel}
\usepackage{float}
\usepackage{balance}
\restylefloat{table}

\usepackage{times}
\usepackage{textcomp}
\usepackage{mathptmx}
\usepackage{courier}
\usepackage{pifont}
\usepackage{amssymb}
\usepackage{eucal}
\usepackage[T1]{fontenc}
\begin{document}




\title{Reuse Detector: Improving the management of STT-RAM SLLCs}
\author{R. Rodr\'iguez-Rodr\'iguez$^1$, J. D\'iaz$^2$, F. Castro$^1$, P. Ib\'a\~nez$^2$, D. Chaver$^1$, V. Vi\~nals$^2$, J.C. Saez$^1$, M. Prieto-Matias$^1$, L. Pi\~nuel$^1$, T. Monreal$^3$, J.M. Llaber\'ia$^3$}
\affiliation{$^1$ArTeCS Group, Facultad de Informatica, University Complutense of Madrid,
$^2$Computer Architecture Group, University of Zaragoza,
$^3$Computer Architecture Department, Technical University of Catalonia}
\email{rrodriguezr@ucm.es, jdmaag@unizar.es, fcastror@ucm.es, imarin@unizar.es, dani02@ucm.es, victor@unizar.es, jcsaezal@ucm.es, mpmatias@ucm.es, lpinuel@ucm.es, teresa@ac.upc.edu, llaberia@ac.upc.edu}

\shortauthors{Rodriguez-Rodriguez et al}


\begin{abstract}
Various constraints of Static Random Access Memory (SRAM) are 
leading to consider new memory technologies as candidates for building on-chip 
shared last-level caches (SLLCs). Spin-Transfer Torque RAM (STT-RAM) is currently 
postulated as the prime contender due to its better energy efficiency, smaller die 
footprint and higher scalability. However, STT-RAM also exhibits some drawbacks, 
like slow and energy-hungry write operations, that need to be mitigated before 
it can be used in SLLCs for the next generation of computers.
In this work we address these shortcomings by leveraging a new management mechanism for STT-RAM SLLCs. 
This approach is based on the previous observation that although the stream of references arriving at the SLLC of a Chip MultiProcessor 
(CMP) exhibits limited temporal locality, it does exhibit \emph{reuse locality}, 
i.e., those blocks referenced several times manifest high probability of forthcoming reuse. 
As such, conventional STT-RAM SLLC management mechanisms, mainly focused on exploiting temporal locality, result in low efficient behavior. 
In this paper, we employ a cache management mechanism that selects the contents of the SLLC 
aimed to exploit reuse locality instead of temporal locality. Specifically, our proposal consists in the inclusion of a 
\emph{Reuse Detector} between private cache levels and the STT-RAM SLLC. Its mission is 
to detect blocks that do not exhibit reuse, in order to avoid their insertion in the SLLC, hence reducing the number of 
write operations and the energy consumption in the STT-RAM. Our evaluation, using multiprogrammed
workloads in quad-core, eight-core and 16-core systems, reveals that our scheme reports on average, energy reductions in the SLLC in the range of 37-30\%, additional energy savings in the main memory in the range of 6-8\% and performance improvements of 3\% (quad-core), 7\% (eight-core) and 14\% (16-core) compared to an STT-RAM SLLC baseline where no reuse detector is employed. More importantly, our approach outperforms DASCA, the state-of-the-art STT-RAM SLLC management, reporting --depending on the specific scenario and the kind of applications used-- SLLC energy savings in the range of 4-11\% higher than those of DASCA, delivering higher performance in the range of 1.5-14\%, and additional improvements in DRAM energy consumption in the range of 2-9\% higher than DASCA. 

\end{abstract}

\maketitle

\section{Introduction}

In the last years chip multiprocessors have become majority on many off-the-shelf systems, such as high performance servers, desktop systems, mobile devices and embedded systems. In all of them the designers usually include a multilevel memory hierarchy, where the shared last-level cache (SLLC) plays an important role in terms of cost, performance and energy consumption. As for the cost, the SLLC generally occupies a chip area similar or even bigger than that of cores. Regarding performance and energy consumption, the SLLC is the last resource before accessing the main memory, which delivers higher energy consumption and lower performance as it is located outside the chip. 

The technologies currently employed in building SLLCs are mainly SRAM or embedded DRAM. However, they both reveal as power-hungry, especially for the large sizes required as the number of cores increases. One way of mitigating this problem is to employ emerging non-volatile memory technologies. Among them, Spin-Transfer Torque RAM (STT-RAM) is clearly the prime contender. STT-RAM removes almost all the static power consumption and, compared to SRAM, provides higher density and therefore much higher capacity within the same budget. Moreover, it delivers higher read efficiency in terms of latency and energy. Nevertheless, some obstacles restrict the adoption of STT-RAM as last-level cache for the next generation of CMPs: its write operation is slower and requires more energy than an SRAM cache. These constraints may lead to a performance drop and even to almost cancel the energy savings derived from the minimal static power consumption of STT-RAM.

In addition, previous research states that conventional SLLC designs are inefficient since they waste most storage space~\cite{dasca_14,deadblock_10}. This is due to the fact that SLLC management policies often lead to store \emph{dead blocks}, i.e., blocks that will not be referenced again before eviction. Indeed, it is frequent that blocks were dead the very first time they enter into the SLLC. This is mainly because the cache levels closer to the processor exploit most of the \emph{temporal locality}, which therefore becomes largely filtered before accessing the SLLC. With the goal of avoiding this effect and hence increasing the hit rate, various mechanisms that modify the SLLC insertion and replacement policies have been proposed recently.

This work addresses the shortcomings aforementioned by focusing on improving the efficiency, in terms of both performance and energy, of a non-inclusive and non-exclusive STT-RAM SLLC in a chip multiprocessor system. Notably, we present a new mechanism of content selection for last-level caches that benefits from the \emph{reuse locality} that SLLC references exhibit~\cite{albericio_taco13,albericio_micro13}. This locality lies in the following principle: when a block is referenced twice in the last level cache (i.e. it is reused), this block has a good chance of being referenced again in the near future. Our approach pursues to insert in the SLLC only those blocks that exhibit reuse at that level. 
For this purpose, we propose to include a new hardware resource between the SLLC and the private cache levels --referred to as \emph{Reuse Detector}-- which determines for each block evicted from the private cache levels if it has been reused or not at the SLLC. If the block is determined to having been reused, it is inserted (or it updates) in the SLLC. Otherwise, the block \emph{bypasses} the SLLC and is sent directly to main memory. 

Our proposal is evaluated in quad, eight and 16-core systems running multiprogrammed workloads, and our experimental results reveal that the reuse detector avoids the insertion of low-utility blocks in the STT-RAM SLLC, making it easier to retain most of reused blocks. This enables us to reduce the amount of the slow and energy-hungry writes to the STT-RAM SLLC, which translates into energy consumption reduction and system performance improvement, outperforming other recent approaches.

The rest of the paper is organized as follows: Section 2 motivates our work and explains some necessary background. 
Section 3 presents our proposal to improve the STT-RAM LLC management. Sections 4 and 5 detail the experimental framework used and the obtained results, respectively. Section 6 recaps some related work and finally, Section 7 concludes the paper.

\section{Background and motivation}
\label{sec:motiv}

In this section we first motivate the need of a new SLLC management scheme by describing the main limitations of SRAM technology and conventional management. Then we briefly describe the DASCA scheme, which is the closest approach to our work and the state-of-the-art STT-RAM SLLC management scheme~\cite{dasca_14}.

\subsection{SRAM and STT-RAM technologies}

As stated above, various emerging technologies are currently considered to replace SRAM as the building-block for SLLCs, being STT-RAM the best placed to overcome SRAM constraints, such as energy consumption and read operation latency. 

\begin{figure}
     \begin{center}
        \subfigure[ ]{%
            \label{fig:stta}
            \includegraphics[width=0.45\textwidth,natwidth=600,natheight=491]{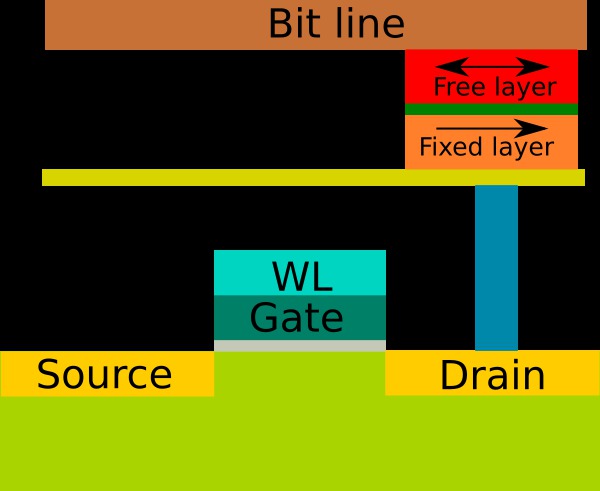}
        }%
        \subfigure[ ]{%
           \label{fig:sttb}
           \includegraphics[width=0.25\textwidth,natwidth=374,natheight=271]{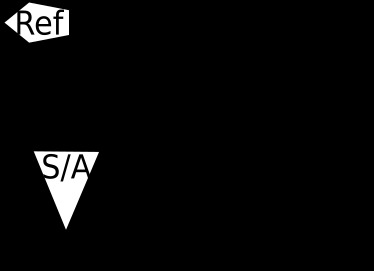}
        }
    \end{center}
    \caption{%
        (a) STT-RAM memory cell structure. (b) STT-RAM equivalent circuit.}%
\end{figure}

The main difference between STT-RAM and SRAM is that the information carrier of the former is a Magnetic Tunnel Junction (MTJ) instead of electric charges. An MTJ contains two ferromagnetic layers (denoted as free and fixed layers) and one tunnel barrier layer, see Figure~\ref{fig:stta}. The fixed layer has a fixed magnetic direction while the free layer can change its magnetic direction by passing a current through the MTJ. If the two ferromagnetic layers have different directions, the MTJ resistance is high, indicating a ``1" state; if the two layers have the same direction, the MTJ resistance is low, indicating a ``0" state. A read operation to an MTJ is performed by applying a small voltage difference between two electrodes of the MTJ and sensing the current flow (see Figure~\ref{fig:sttb}, where the STT-RAM cell is represented as a variable resistor). A write operation is performed by applying a large voltage difference between two electrodes for a given duration called write pulse width.



Table~\ref{tab:param} shows the key features of an 1-bank 1MB LLC implemented with SRAM and STT-RAM 22 nm technology, modeled with CACTI 6.5~\cite{cacti65} and NVSim~\cite{nvsim_12}, respectively. As shown, an STT-RAM cache exhibits smaller die footprint and better efficiency in read operations than an SRAM cache. More importantly, an STT-RAM cache consumes almost two orders of magnitude less static power compared to SRAM. Conversely, the STT-RAM cache exhibits a significant drawback that needs to be mitigated: the poor write performance both in terms of latency and energy consumption. 

\begin{table}
\begin{center}
\scriptsize
\tbl{Area, latency and energy consumption for 22 nm SRAM and STT-RAM 1MB caches.\label{tab:param}}{ 
    \resizebox{1\textwidth}{!}{%
\begin{tabular}{|l|c|c|c|}
\hline
\textbf{Parameter}&\textbf{SRAM}&\textbf{STT-RAM}&\textbf{Ratio SRAM/STT-RAM}\\
\hline
Area ($mm^2$) & 0.94 & 0.35 & 2.68\\
\hline
Read Latency (ns) & 8.75 & 5.61 & 1.56\\
\hline
Write Latency (ns) & 8.75 & 16.5 & 0.53\\
\hline
Read Energy (nJ) & 0.56 & 0.32 & 1.75\\
\hline
Write Energy (nJ) & 0.56 & 1.31 & 0.43\\
\hline
Leakage Power (mW) & 190.58 & 3.09 & 61.67\\
\hline
\end{tabular}}}
\end{center}
\end{table}

\subsection{SLLC management techniques}

Regardless of implementation technology, last-level caches usually suffer from the same problem: they keep data assuming that recently referenced lines are likely to appear in the near future (\emph{temporal locality}). Nevertheless, various recent studies point out that the reference stream entered in the SLLC does not usually exhibit temporal locality. Notably, in~\cite{albericio_taco13} the authors observe that this reference stream exhibits \emph{reuse locality} instead of temporal locality. Essentially, that term describes the property that the second reference to a line is a good indicator of forthcoming reuse and also that recently reused lines are more valuable than other lines reused longer.

\begin{figure}[htb]
\resizebox{1.01\linewidth}{!}{\includegraphics{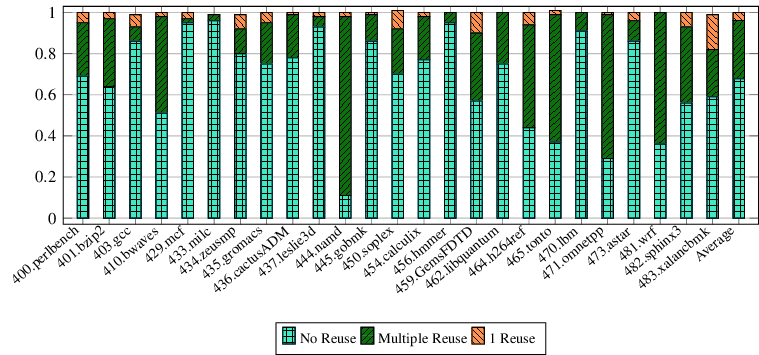}}
\caption{Breakdown of blocks replaced from the LLC according to the number of accesses they receive before eviction.}
\label{motiv_reuso}
\end{figure}

The studies carried out in ~\cite{albericio_taco13,albericio_micro13,diaz_jp15} demonstrate, considering a large amount of multiprogrammed workloads and different multiprocessor configurations, two important aspects: first, most lines in the SLLC are dead (they will not receive any further hits during their lifetime in the SLLC) and second, most SLLC hits come from a small subset of lines. We have performed our own analysis about the behavior of the blocks evicted from the SLLC, in terms of the amount of accesses they receive before eviction, in the scenario 
with the configuration detailed in Table~\ref{tab:cpuSpecs}, but, as a starting point, employing just one core. 
Figure~\ref{motiv_reuso} illustrates this behavior, grouping the blocks into three different categories: no reuse, just one reuse or more than one reuse (multiple reuse). 

As shown, our experimental results confirm that most lines in the LLC are dead. Notably, around 70\% of the blocks do not receive any further access since the moment they enter the LLC. Only around 5\% of the blocks just receives one further hit (i.e. one reuse) and around 25\% exhibit more than one reuse. 
 


Consequently, getting blocks with just one use (the block fill, so no reuse) to bypass the LLC when they are evicted from the previous level caches, and just storing blocks with reuse (at least two LLC accesses), should allow to hold the small fraction of blocks with multiple reuses, increasing the LLC hit rate and improving system performance.

\begin{figure}[htb]
\resizebox{1.01\linewidth}{!}{\includegraphics{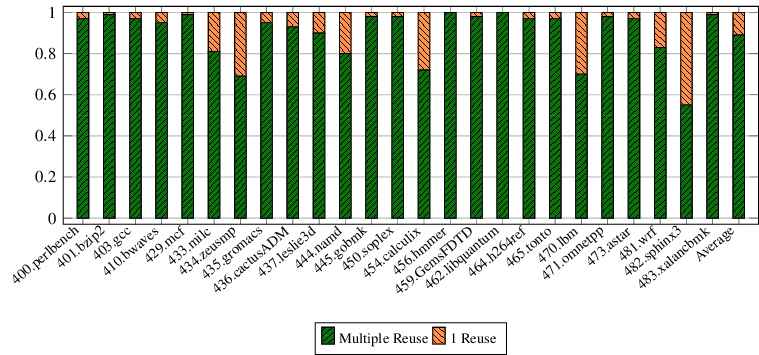}}
\caption{Breakdown of block hits at the LLC according to the number of accesses they have received before read.}
\label{motiv_Hitreuso}
\end{figure}

Furthermore, Figure~\ref{motiv_Hitreuso} shows that most LLC hits are to blocks having multiple reuses, which together with the aforementioned fact that most blocks inserted in the LLC do not experience any reuse, highly justify the idea of a content selector based on reuse detection between private caches and LLC.

\subsection{DASCA scheme}

In~\cite{dasca_14}, the authors propose \emph{Dead Write Prediction Assisted STT-RAM Cache Architecture} (DASCA) to predict and bypass dead writes (writes to data in last level caches not referenced again during the lifetime of corresponding cache blocks) for write energy reduction. In this work dead writes are classified into three categories: dead-on-arrival fills, dead-value fills and closing writes, as a theoretical model for redundant write elimination. On top of that they also present a dead write predictor based on a state-of-the-art dead block predictor~\cite{deadblock_10}. Thus, DASCA bypasses a write operation to the SLLC only if it is predicted not to incur extra cache misses. 

Notably, DASCA adds a specific field to each line at the private levels to store the PC (program counter) of the instructions writing a block, being this PC only updated upon write operations. Also, a PC-signature table (prediction table) is included in the design in order to make the prediction about dead writes (this table is updated according to the mechanism shown in the Table 2 of the paper itself~\cite{dasca_14}). Specifically, the mechanism samples a few cache sets and keeps track of PC information only for those sets. Predictions are made via the predictor table, made up of saturating counters similar to those used in a bimodal branch predictor, being the counters indexed by the signatures stored in the sampler entries. Thus, this PC-based predictor correlates dead blocks with addresses of memory instructions (signatures), so that different signatures are used depending on the kind of dead write predicted. 

\section{Design}
\label{sec:des}

In this section we first describe the baseline system we start from. Then we describe in detail the proposed design built on top of that. 



\subsection{Baseline system}


The memory hierarchy used in the baseline multi-core system includes two private levels (L1 and L2) and a last-level cache shared among all the cores (SLLC). All caches are write-back, write-allocate and LRU. L1 and L2 are inclusive while the SLLC is non inclusive. 

The baseline management of this memory hierarchy is as follows: When a block is requested to Main Memory (MM), it is copied to the private cache levels of the requester core, but not to the SLLC. During its lifetime at the private levels of the core, the block can be requested by other cores, in which case a copy will be sent from a private L2 cache to the L1-L2 caches of the requesting core, as dictated by the directory-based coherency mechanism (please refer to Section~\ref{sec:sim} for more details on the coherency mechanism).

When a block is evicted from an L2 cache, the SLLC is checked: In case the block is not present there (either because it has not been inserted yet or because it has already been inserted and evicted by the SLLC replacement mechanism), it is inserted in the SLLC, otherwise if the block is already in the SLLC, the block is updated or just discarded, depending respectively on whether the block is dirty or clean. Thus, in our hierarchy, SLLC insertions never come from MM but from an L2, in a similar way to an exclusive policy. Note however that our mechanism differs from an exclusive policy in that, as a result of a hit in the SLLC, the block is copied to the private cache levels of the requester core, but maintained in the SLLC.

\subsection{The Reuse Detector}


As explained earlier, several works have demonstrated that a notable percentage of the blocks inserted/updated in the SLLC are in fact useless, as they are dead-blocks (i.e. blocks which will not be referenced any more during their lifetime in the SLLC)~\cite{dasca_14,deadblock_10}. These useless blocks are harmful, as they evict other blocks which could potentially be useful in the future, and moreover, they increase the amount of writes to the SLLC, which in the context of NVMs (Non-Volatile Memories) is far from convenient, as explained in previous sections.

\begin{figure}[htb]
\includegraphics{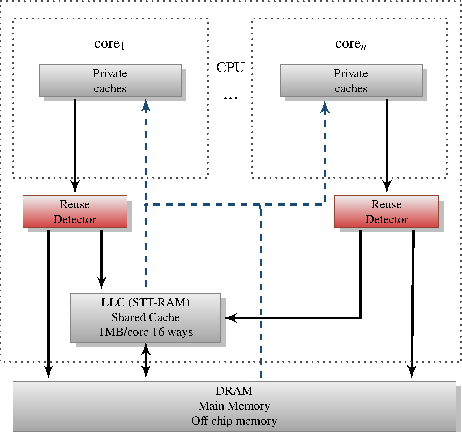}	
\caption{Placement of a Reuse Detector between each private L2 level and STT-RAM SLLC.}
\label{system2Simulate}
\end{figure}

In this paper we leverage the technique for reducing the amount of dead-blocks inserted/updated in the SLLC~\cite{diaz_jp15} to improve the efficiency of a STT-RAM SLLC. 
In~\cite{diaz_jp15}, the authors present a proposal that, in an exclusive memory hierarchy, reduces the amount of blocks inserted in a conventional SLLC by around 90\%. We apply this technique to a different context, i.e., to a non-inclusive STT-RAM SLLC design within a memory hierarchy where L1-L2 are inclusive. The exclusion policy employed in~\cite{diaz_jp15} implies that, upon a SLLC hit, the block is copied to the private cache levels and removed from the SLLC. In our case, the block is inserted in the SLLC at the end of its usage in the private caches and remains in the SLLC until eviction.
For our purpose, we include an intermediate element between each private L2 cache and the SLLC (Figure~\ref{system2Simulate}). A block evicted from the private L2 caches is targeted to the corresponding element, which we denote as Reuse Detector (RD), instead of accessing directly to the SLLC as it would do in the baseline system. The RD decides whether to send the block to the SLLC or not (i.e. to bypass the shared last-level cache), by means of a prediction about the future usefulness of the block. We must highlight that, being the RD out of the block request path to the SLLC, it does not impact the SLLC hit or miss latencies.

For accomplishing the RD prediction, we apply Albericio\textquotesingle s concept of reuse locality~\cite{albericio_taco13,albericio_micro13}. As such, if the evicted block from the L2 has never been requested to the SLLC since the time it entered the cache hierarchy (i.e. it has never been reused at the SLLC), the block is predicted as a dead-block and thus it bypasses the SLLC, directly updating MM (if the block is dirty) or being just discarded (if it is clean). Otherwise, if the block has been reused (i.e. it has been requested to the SLLC at least once since the time it entered the cache hierarchy) and thus it is predicted as a non-dead-block, it is inserted/updated in the SLLC.

The RD consists of a FIFO buffer and some management logic. The buffer stores the tag addresses of the blocks evicted by the private levels in order to maintain their reuse state. Moreover, an extra bit, called \emph{reuse bit}, is added to each cache line in the private levels. This bit distinguishes if the block was inserted at the private cache levels from the SLLC or from another private cache level (\emph{reuse} bit is 1), or main memory (\emph{reuse} bit is 0). In the following sections, we analyze in detail the Reuse Detector operation and implementation.

\subsubsection{Reuse Detector operation}

As we said in a previous section, our proposal aims to reduce the amount of writes to the STT-RAM SLLC and to improve the management of SLLC blocks, which translate into system performance improvement and energy consumption reduction on the system.

\begin{figure}[htb]
\resizebox{\linewidth}{!}{\includegraphics{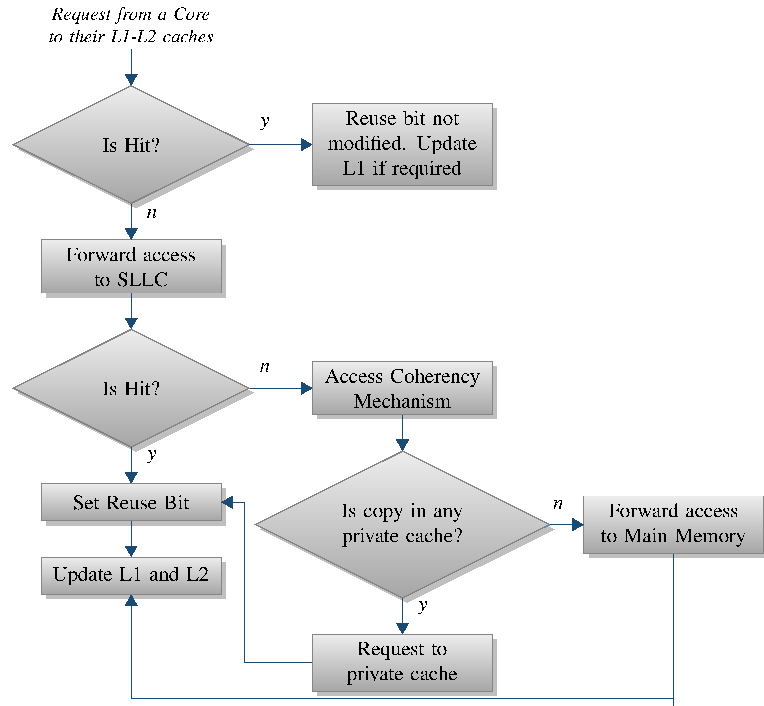}}
\caption{Block request and reuse bit management.}
\label{fig:flow1}
\end{figure}

Figure~\ref{fig:flow1} shows a flow diagram of a block request from a core to its associated private caches. If the request hits in L1 or L2 the reuse bit is untouched, and the block is copied in L1 if it was not there (inclusive policy). Otherwise, the request is forwarded to the SLLC. If the access hits in the SLLC, the block is provided to the core and copied in the private levels with the reuse bit set. If the access misses in the SLLC but the coherency mechanism informs that the block is present in another private cache, the block is provided by that cache. In this case, the access is recognized as a reuse, so the reuse bits are also set. Finally, if no copy of the block is present in the cache hierarchy, it is requested to MM and copied in L1-L2 with the reuse bit unset.

\begin{figure}[htb]
\resizebox{\linewidth}{!}{\includegraphics{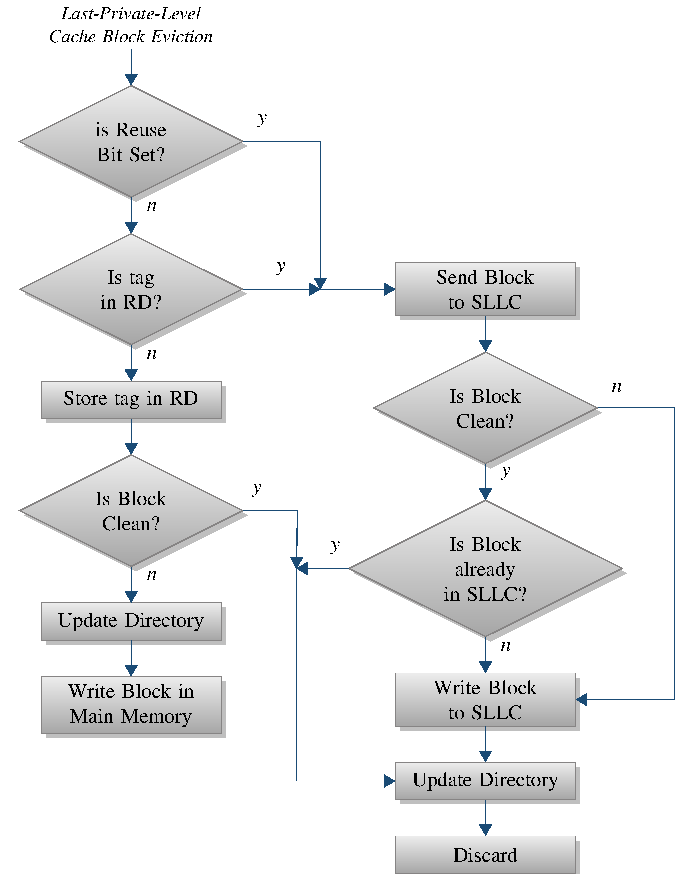}}
\caption{Block eviction from a private cache and SLLC insertion.}
\label{fig:flow2}
\end{figure}

Figure~\ref{fig:flow2} shows a flow diagram of a block eviction from an L2 cache (if required, the corresponding L1 cache is invalidated). When a block is evicted from a last private level cache, its reuse bit is checked. If the reuse bit is set, it means that the block was inserted into the private caches either from the SLLC or from another private cache after checking the SLLC and the coherency mechanism. In any case, the block is considered as having been reused, and it should be inserted in the SLLC (if not present yet) or just updated (if the block is dirty but it is already present in the SLLC). Note that if the block is clean and already present in the SLLC, it can just be discarded. If the reuse bit is unset (i.e. the block was brought into the private caches directly from main memory) but the block\textquotesingle s tag is found in the RD buffer, the block is also considered as having been reused, and thus it is handled as in the previous situation. Finally, if the reuse bit is unset and its tag is not present in the RD buffer, it means that the block is considered as not having been reused yet. Based again on Albericio\textquotesingle s observations~\cite{albericio_taco13,albericio_micro13}, the block should bypass the SLLC, as it is predicted as a dead-block, and it should be sent to MM (if the block is dirty) or just discarded (if it is clean). Note that in all cases the coherency mechanism must be updated.

\subsubsection{Example}
\begin{figure*}
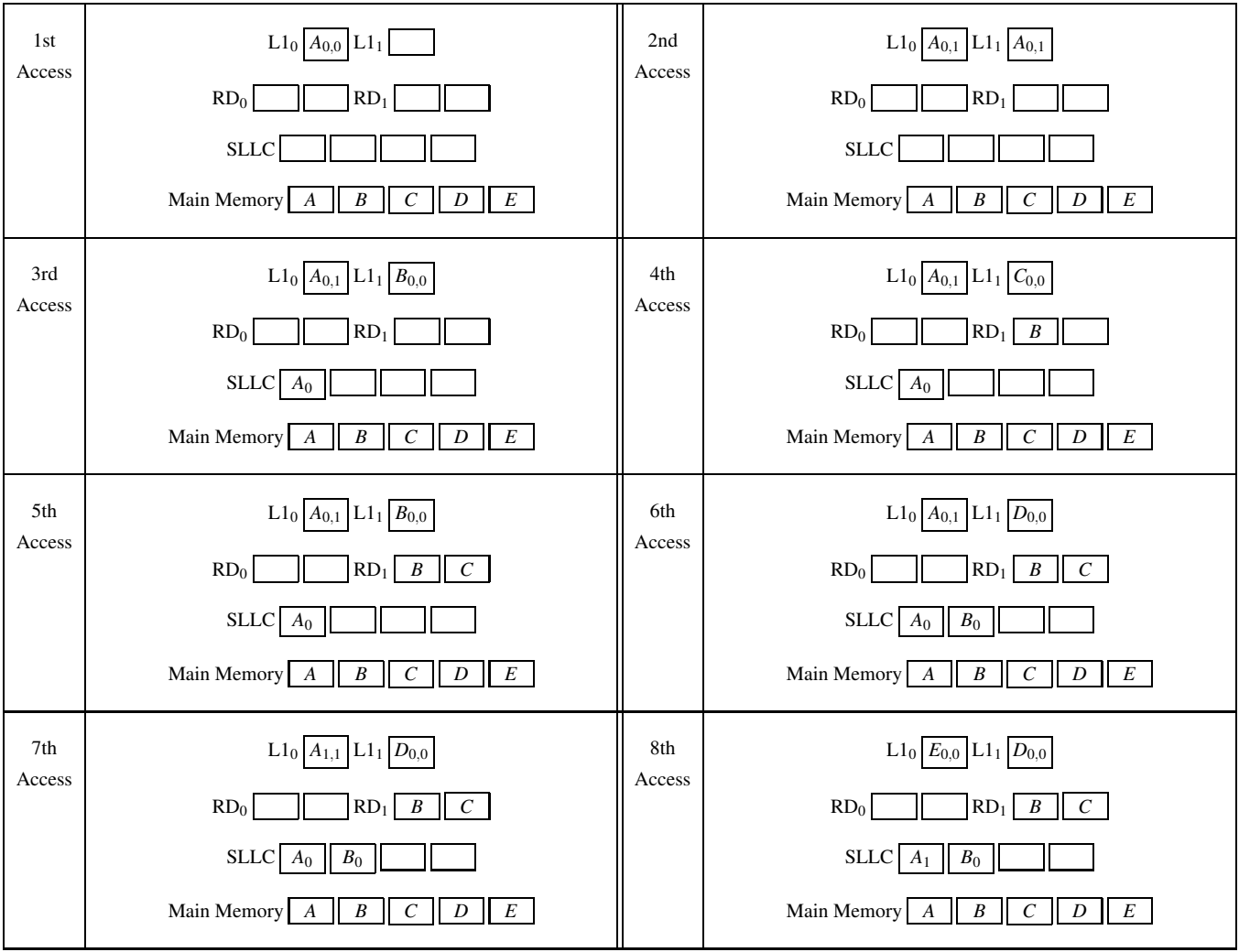

\begin{center}
\resizebox{1\textwidth}{!}{%
\begin{tabular}[h]{|c|C{9cm}||c|C{9cm}|}


%
%



  \hline 

  & & &\\
  1st& L1$_0$ $\framebox[0.8cm]{$A_{0,0}$}$ L1$_1$ $\framebox[0.8cm]{$\phantom{X}$}$ & 2nd& L1$_0$ $\framebox[0.8cm]{$A_{0,1}$}$ L1$_1$ $\framebox[0.8cm]{$A_{0,1}$}$\\
  Access& &Access &\\
  & RD$_0$ $\framebox[0.8cm]{$\phantom{X}$}$ $\framebox[0.8cm]{$\phantom{X}$}$ RD$_1$ $\framebox[0.8cm]{$\phantom{X}$}$ $\framebox[0.8cm]{$\phantom{X}$}$ & & RD$_0$ $\framebox[0.8cm]{$\phantom{X}$}$ $\framebox[0.8cm]{$\phantom{X}$}$ RD$_1$ $\framebox[0.8cm]{$\phantom{X}$}$ $\framebox[0.8cm]{$\phantom{X}$}$\\
  & & &\\
  & SLLC $\framebox[0.8cm]{$\phantom{X}$}$ $\framebox[0.8cm]{$\phantom{X}$}$ $\framebox[0.8cm]{$\phantom{X}$}$ $\framebox[0.8cm]{$\phantom{X}$}$ & & SLLC $\framebox[0.8cm]{$\phantom{X}$}$ $\framebox[0.8cm]{$\phantom{X}$}$ $\framebox[0.8cm]{$\phantom{X}$}$ $\framebox[0.8cm]{$\phantom{X}$}$ \\  
  & & &\\
  & Main Memory $\framebox[0.8cm]{$A$}$ $\framebox[0.8cm]{$B$}$ $\framebox[0.8cm]{$C$}$ $\framebox[0.8cm]{$D$}$ $\framebox[0.8cm]{$E$}$  & & Main Memory $\framebox[0.8cm]{$A$}$ $\framebox[0.8cm]{$B$}$ $\framebox[0.8cm]{$C$}$ $\framebox[0.8cm]{$D$}$ $\framebox[0.8cm]{$E$}$ \\  
  & & &\\ \hline


  & & &\\ 
  3rd & L1$_0$ $\framebox[0.8cm]{$A_{0,1}$}$ L1$_1$ $\framebox[0.8cm]{$B_{0,0}$}$ & 4th & L1$_0$ $\framebox[0.8cm]{$A_{0,1}$}$ L1$_1$ $\framebox[0.8cm]{$C_{0,0}$}$\\
  Access& &Access &\\
  & RD$_0$ $\framebox[0.8cm]{$\phantom{X}$}$ $\framebox[0.8cm]{$\phantom{X}$}$ RD$_1$ $\framebox[0.8cm]{$\phantom{X}$}$ $\framebox[0.8cm]{$\phantom{X}$}$ & & RD$_0$ $\framebox[0.8cm]{$\phantom{X}$}$ $\framebox[0.8cm]{$\phantom{X}$}$ RD$_1$ $\framebox[0.8cm]{$B$}$ $\framebox[0.8cm]{$\phantom{X}$}$\\
  & & &\\
  & SLLC $\framebox[0.8cm]{$A_0$}$ $\framebox[0.8cm]{$\phantom{X}$}$ $\framebox[0.8cm]{$\phantom{X}$}$ $\framebox[0.8cm]{$\phantom{X}$}$ & & SLLC $\framebox[0.8cm]{$A_0$}$ $\framebox[0.8cm]{$\phantom{X}$}$ $\framebox[0.8cm]{$\phantom{X}$}$ $\framebox[0.8cm]{$\phantom{X}$}$\\  
  & & &\\
  & Main Memory $\framebox[0.8cm]{$A$}$ $\framebox[0.8cm]{$B$}$ $\framebox[0.8cm]{$C$}$ $\framebox[0.8cm]{$D$}$ $\framebox[0.8cm]{$E$}$  & & Main Memory $\framebox[0.8cm]{$A$}$ $\framebox[0.8cm]{$B$}$ $\framebox[0.8cm]{$C$}$ $\framebox[0.8cm]{$D$}$ $\framebox[0.8cm]{$E$}$ \\  
  & & &\\ \hline


  & & &\\ 
  5th& L1$_0$ $\framebox[0.8cm]{$A_{0,1}$}$ L1$_1$ $\framebox[0.8cm]{$B_{0,0}$}$ & 6th& L1$_0$ $\framebox[0.8cm]{$A_{0,1}$}$ L1$_1$ $\framebox[0.8cm]{$D_{0,0}$}$\\
  Access& &Access &\\
  & RD$_0$ $\framebox[0.8cm]{$\phantom{X}$}$ $\framebox[0.8cm]{$\phantom{X}$}$ RD$_1$ $\framebox[0.8cm]{$B$}$ $\framebox[0.8cm]{$C$}$ & & RD$_0$ $\framebox[0.8cm]{$\phantom{X}$}$ $\framebox[0.8cm]{$\phantom{X}$}$ RD$_1$ $\framebox[0.8cm]{$B$}$ $\framebox[0.8cm]{$C$}$\\
  & & &\\
  & SLLC $\framebox[0.8cm]{$A_0$}$ $\framebox[0.8cm]{$\phantom{X}$}$ $\framebox[0.8cm]{$\phantom{X}$}$ $\framebox[0.8cm]{$\phantom{X}$}$ &  & SLLC $\framebox[0.8cm]{$A_0$}$ $\framebox[0.8cm]{$B_0$}$ $\framebox[0.8cm]{$\phantom{X}$}$ $\framebox[0.8cm]{$\phantom{X}$}$\\  
  & & &\\
  & Main Memory $\framebox[0.8cm]{$A$}$ $\framebox[0.8cm]{$B$}$ $\framebox[0.8cm]{$C$}$ $\framebox[0.8cm]{$D$}$ $\framebox[0.8cm]{$E$}$  &  & Main Memory $\framebox[0.8cm]{$A$}$ $\framebox[0.8cm]{$B$}$ $\framebox[0.8cm]{$C$}$ $\framebox[0.8cm]{$D$}$ $\framebox[0.8cm]{$E$}$ \\  
  & & &\\ \hline


  & & &\\ 
  7th& L1$_0$ $\framebox[0.8cm]{$A_{1,1}$}$ L1$_1$ $\framebox[0.8cm]{$D_{0,0}$}$ & 8th & L1$_0$ $\framebox[0.8cm]{$E_{0,0}$}$ L1$_1$ $\framebox[0.8cm]{$D_{0,0}$}$\\
  Access& &Access &\\
  & RD$_0$ $\framebox[0.8cm]{$\phantom{X}$}$ $\framebox[0.8cm]{$\phantom{X}$}$ RD$_1$ $\framebox[0.8cm]{$B$}$ $\framebox[0.8cm]{$C$}$ & & RD$_0$ $\framebox[0.8cm]{$\phantom{X}$}$ $\framebox[0.8cm]{$\phantom{X}$}$ RD$_1$ $\framebox[0.8cm]{$B$}$ $\framebox[0.8cm]{$C$}$ \\
  & & &\\
  & SLLC $\framebox[0.8cm]{$A_0$}$ $\framebox[0.8cm]{$B_0$}$ $\framebox[0.8cm]{$\phantom{X}$}$ $\framebox[0.8cm]{$\phantom{X}$}$ & & SLLC $\framebox[0.8cm]{$A_1$}$ $\framebox[0.8cm]{$B_0$}$ $\framebox[0.8cm]{$\phantom{X}$}$ $\framebox[0.8cm]{$\phantom{X}$}$ \\  
  & & &\\
  & Main Memory $\framebox[0.8cm]{$A$}$ $\framebox[0.8cm]{$B$}$ $\framebox[0.8cm]{$C$}$ $\framebox[0.8cm]{$D$}$ $\framebox[0.8cm]{$E$}$ & & Main Memory $\framebox[0.8cm]{$A$}$ $\framebox[0.8cm]{$B$}$ $\framebox[0.8cm]{$C$}$ $\framebox[0.8cm]{$D$}$ $\framebox[0.8cm]{$E$}$  \\   
  & & &\\ \hline
\end{tabular}}
\caption{Example of the Reuse Detector operation.}
\label{tableExampleReuseDet}
\end{center}
\end{figure*}

For the sake of clarifying the Reuse Detector operation, in this subsection we provide a straightforward example illustrating the flow of five memory blocks (A, B, C, D and E) through the different cache levels under a given access pattern. In this example, we consider a dual-core system ($Core_{0}$ and $Core_{1}$) with private first level caches ($L1_{0}$ and $L1_{1}$), a shared second level cache (SLLC), and the corresponding Reuse Detectors between both cache levels. In the example we assume a simplified configuration where: 1) direct-mapped L1s, 2-way set associative RDs and 4-way set associative SLLC are considered; 2) all memory blocks map to the same L1 frame and to the same RD and SLLC set; and 3) initially, all caches and RDs are empty. Next, we detail the access sequence of our example and show the contents of the memory hierarchy after each access in Figure~\ref{tableExampleReuseDet}. Note that we specify as a subindex the dirty bit followed by the reuse bit ($X_{d,r}$) for each block \emph{X} in the private cache levels, and only the dirty bit ($X_d$) for each block \emph{X} in the SLLC.  

\begin{enumerate}
\item $Core_0$ requests a word within block A for reading: The access misses in $L1_0$, it is forwarded to the SLLC, and given that the access to SLLC also misses and the block is not present in any other private cache, it is forwarded to MM. According to Figure~\ref{fig:flow1}, block A is copied to $L1_0$ with its reuse bit unset, and the requested word is provided to $Core_0$. 
\item $Core_1$ requests a word within block A for reading: The access misses in $L1_1$ and SLLC. However, the coherency mechanism informs that the block is at $L1_0$, so the request is forwarded to that cache. According to Figure~\ref{fig:flow1}, the block is copied to $L1_1$ and both reuse bits are set, as we recognize this access as an SLLC reuse. 
\item $Core_1$ requests a word within block B for reading: The access misses in $L1_1$ and SLLC, and the block is not present in any other private cache, so the request is forwarded to MM. According to Figure~\ref{fig:flow1}, block B is copied to $L1_1$ (replacing block A) with its reuse bit unset, and the requested word is provided to $Core_1$. According to Figure~\ref{fig:flow2}, given that block A had its reuse bit set, it is inserted into the SLLC. 
\item $Core_1$ requests a word within block C for reading: Block C is inserted in $L1_1$ and block B is replaced. As the reuse bit of block B was unset and its tag was not in $RD_1$, according to Figure~\ref{fig:flow2} the tag is stored in $RD_1$ and, given that the block is clean, it is not inserted in the SLLC but just discarded. 
\item $Core_1$ requests a word within block B for reading: This access is handled analogously to the previous access. 
\item $Core_1$ requests a word within block D for reading: Block D is inserted in $L1_1$ and block B is replaced. As the reuse bit of block B was unset but its tag was present in $RD_1$, according to Figure~\ref{fig:flow2} block B is inserted in the SLLC.
\item $Core_0$ writes to a word within block A: The access hits in $L1_0$. The dirty bit for the block is set.
\item $Core_0$ requests a word within block E for reading: Block E is inserted in $L1_0$ and block A is replaced. As the dirty bit of block A is set and A is already present in the SLLC, the block is updated at this level.
\end{enumerate}

\subsubsection{Implementation details}
\label{subsubsec:details}

Although a typical set-associative design could be used for the RD implementation, where a whole block tag, a validity bit and some information related with the replacement policy is included for each line, as in~\cite{diaz_jp15} we use two techniques aimed at reducing the required space: sectoring and compression.
A sector is a set of consecutive memory blocks aligned to the sector size. Storing sector tags in the RD allows to merge in a single line of the RD the information related with several blocks. Note that for each entry it is necessary to store a presence bit. For example, with a sector comprising 4-blocks, each line is made up of a tag derived from the sector address, a validity bit, some bits storing the replacement state and 4 presence bits.

The compression of the tags is achieved based on the following process: let \emph{t} be the amount of bits of the full tag and \emph{c} the amount of bits of the compressed tag, being t$>$c. We first divide the full tag into several pieces, each of size \emph{c} (the last piece is filled with \emph{0s} if necessary). Then, we \emph{xor} all the pieces, obtaining the compressed tag. Note that each compressed tag is shared among various sectors, thus false positives are possible where non-reused blocks are delivered to the SLLC. This situation does not cause a functional problem, but it may degrade system performance, so the value of \emph{c} must be chosen carefully.

\begin{figure}[htb]
\includegraphics{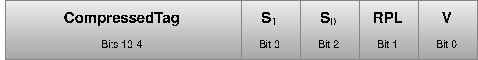}	
\caption{Reuse Detector entry.}
\label{fig:rd_block}
\end{figure}

As for the storage overhead of the RD implementation, i.e., the total amount of extra bits required compared to the baseline, we need the following hardware: The RD has 1024 sets and 16 ways (our simulations reveal that this value provides similar performance to that of higher associativity values), and a sector size of 2 blocks. Each RD entry requires 14 bits (10 for the compressed tag, 2 for the block presence, 1 for the replacement policy and 1 validity bit) as Figure~\ref{fig:rd_block} illustrates. Given that the amount of entries in the RD is 8K, the total extra storage required per core is 14 KB, which represents a negligible 1.3\% of an 1MB last level cache. 


Finally, as RD replacement policy we use a 1-bit FIFO. Again based on our simulations, this scheme delivers a similar performance as other policies that would require more storage. In a FIFO policy, age information is updated only when a new address is inserted, and not during subsequent hits. This approach is fully consistent with the main RD buffer goal of detecting the first reuse of a block.

\subsection{Reuse Detector vs DASCA}

In this section we briefly discuss the main differences between the RD approach and the DASCA scheme, which will be further extended and analyzed in the evaluation section. As for the operation of both approaches, note that the DASCA mechanism tries to predict dead writes based on a PC-based predictor. For this purpose, the PC signature of each block that accesses the SLLC must be recorded. Conversely, the RD scheme tries to predict dead-blocks based on their reuse state. Our prediction is based on the addresses of the accessed data instead of the instruction addresses used in DASCA. Also, in our approach we directly store the mentioned addresses while in the DASCA scheme the authors employ a PC-signatures table which is trained by an auxiliary cache that works in parallel with the conventional cache.

Focusing on the specific implementation of the DASCA scheme for the evaluation of this proposal, it is worthy to note that our approach employs a memory hierarchy where L1 and L2 are inclusive while the SLLC (L3) is non inclusive, whereas the original DASCA scheme is evaluated in~\cite{dasca_14} employing a memory hierarchy with just two cache levels and assuming non-inclusive caches by default, although the authors also propose a bypassing scheme that supports both inclusive and non-inclusive caches. Therefore, and looking for a fair comparison between RD and DASCA, we implement DASCA using exactly the same three-level non-inclusive non-exclusive hierarchy employed in the RD approach. As a result of that, the only high-level change with respect to the original version of DASCA lies in that one of the three possible cases they use to classify the dead writes (``dead-value fills'', blocks that receive a writeback request from lower-level caches right after the block is filled, but before any read operation, i.e., the filled block data are overwritten before being read), can not exist since they are removed by the inclusion mechanism we employ in our approach. Notably, this is due to the fact that, in our configuration, all the insertions in the SLLC are motivated by an L2 block eviction, and, as L1 and L2 are inclusive, these evicted blocks are only located in the SLLC after the eviction from L2. If after that the processor generates a write request on one of these blocks, a write hit occurs, and consequently the block is copied to the private cache levels and therefore this block in the SLLC can not be written again in the SLLC before being read. Hence, we consider this evaluation as fair, since this way we are evaluating DASCA under the same conditions as our approach, so that we are not giving an advantage to our RD by the fact that it directly avoids the ``dead-value fills'' with the inclusiveness management. 

\section{Experimental Framework}
\label{sec:sim}

\begin{table*}
	\begin{center}
		\scriptsize
		\tbl{CPU and Memory Hierarchy specification. \label{tab:cpuSpecs}}{ 
		\resizebox{0.9\textwidth}{!}{%
		\begin{tabular}{|l|p{42em}|}
			\hline
			\textbf{Architecture}&\textbf{x86}\\
			\hline
				CPUS & 1/4/8, 2GHz\\
				\hline
				Pipeline &  8 Fetch, 8 Decode, 8 Rename, 8 Issue/Execute/Writeback, 8 Commit\\
				\hline
				Registers & Integer (256), Floating Point (256)\\
				\hline
				Buffers & Reorder Buffer (192), Instruction Queue (64)\\
				\hline
				Branch Predictor & TournamentBP \\
				\hline
				Functional Units & IntALU=6, IntMulDiv=2, FPALU=4, FPMultDiv=2, SIMD-Unit=4, RdWrPort=4, IprPort=1\\ 
				\hline
				Private Cache L1 D/I & 32 KB, 8 ways, LRU replacement, Block Size 64B, Access Latency 2 cycles, SRAM\\
				\hline
				Private Cache L2 D/I & 256 KB, 16 ways, LRU replacement, Block Size 64B, Access Latency 5 cycles, SRAM\\
				\hline
				Interconnection & Crossbar network, modeled using Garnet, latency 3 cycles \\
				\hline
				Shared Cache L3 & 1 bank/1MB/core, 16 ways, LRU replacement, Block Size 64B, R/W Latency 6/17 cycles, STT-RAM\\
				\hline
				DRAM & 2 Ranks, 8 Banks, 4kB Page Size, DDR3 1066MHz \\
				\hline
				DRAM Bus & 2 channels with a 8 bus of 8 bits \\
				\hline
		\end{tabular}}}
	\end{center}
\end{table*}

\begin{table*}
	\begin{center}
		\scriptsize
		\tbl{Benchmark characterization according to the number of LLC writes per Kinstruction (WPKI).\label{tab:writeperinst}}{ 
		\resizebox{0.90\textwidth}{!}{%
		\begin{tabular}{|c|c|c|}
			\hline
			\textbf{High}&\textbf{Medium}&\textbf{Low}\\
			\hline
			lbm, mcf, libquantum, bwaves, & bzip2, soplex, gcc, wrf, astar, & gromacs, calculix, h264ref, tonto, \\
			milc, cactusADM, zeusmp, leslie3d & hmmer, xalancbmk, gobmk, perlbench & omnetpp, namd, sphinx3, GemsFDTD \\
			\hline
		\end{tabular}}}
	\end{center}
\end{table*}

For our experiments we use the \emph{gem5} simulator~\cite{Binkert2011} and we employ the \emph{ruby} memory model, specifically the MOESI\_CMP-directory coherence policy provided by the simulator. It is worth noting that we focus on a MOESI policy since protocols with owned state (e.g. MOESI and MOSI) are able to reduce the number of writes to the LLC, as demonstrated by Chang et.al.~\cite{changimpact}. We modify the coherence protocol, encoding the proposed reuse detector. We simulate both a single and a multi-core scenario. For the sake of a better accuracy in both execution modes, an O3 processor type (detailed mode of simulation) was used. 


\begin{table*}
	\begin{center}
		{\fontsize{7}{7}\selectfont
		\caption{SPEC 2006 multiprogrammed mixes for the 4-core CMP.\label{tab:mix4C}}
		\begin{tabular}{|c|c|c|c|}
			\hline
			\textbf{Mixes}&\textbf{Applications}&\textbf{Mixes}&\textbf{Applications}\\
			\hline
			mix.H0 & cactusADM, leslie3d, libquantum, mcf  &mix.H1 & bwaves, cactusADM, leslie3d, milc \\
			\hline
			mix.H2 & libquantum, mcf, milc, zeusmp  &mix.H3 & bwaves, cactusADM, lbm, leslie3d \\
			\hline
			mix.M0 & bzip2, gobmk, soplex, xalancbmk  &mix.M1 & gcc, perlbench, wrf, xalancbmk \\
			\hline
			mix.M2 & gcc, gobmk, hmmer, soplex  &mix.M3 & astar, gobmk, perlbench, wrf \\
			\hline
			mix.L0 & calculix, GemsFDTD, namd, sphinx3 &mix.L1 & gromacs, h264ref, omnetpp, tonto \\
			\hline
			mix.L2 & calculix, GemsFDTD, omnetpp, sphinx3 &mix.L3 & gromacs, h264ref, namd, tonto   \\
			\hline
			mix.HM0 & milc, zeusmp; astar, gcc &mix.HM1 & lbm, leslie3d, libquantum; gobmk \\
			\hline
			mix.HM2 & milc, zeusmp; gcc, gobmk &mix.HM3 & bwaves, mcf; soplex, xalancbmk \\
			\hline
			mix.HL0 & bwaves, cactusADM, leslie3d; omnetpp &mix.HL1 & bwaves, lbm, libquantum; omnetpp \\
			\hline
			mix.HL2 & lbm, leslie3d; gromacs, namd  &mix.HL3 &  leslie3d, milc; GemsFDTD, omnetpp \\
			\hline
			mix.ML0 & perlbench; gromacs, sphinx3, tonto &mix.ML1 & hmmer, wrf; gromacs,h264ref \\
			\hline
			mix.ML2 & perlbench, wrf; GemsFDTD, namd &mix.ML3 & soplex, xalancbmk; sphinx3, tonto \\
			\hline
			mix.HML0 & mcf; hmmer; h264ref, omnetpp &mix.HML1 & milc; hmmer; GemsFDTD, h264ref \\
			\hline
			mix.HML2 & milc; bzip2, wrf; GemsFDTD &mix.HML3 & bwaves, leslie3d; xalancbmk; GemsFDTD,  \\
			\hline
		\end{tabular}}
	\end{center}
\end{table*}

The main features of both the processor and the memory hierarchy are shown in Table~\ref{tab:cpuSpecs}. The network used is a crossbar modeled with Garnet~\cite{agarwal2009garnet}, a detailed interconnection network model inside gem5. As explained above, for the evaluation of our proposed RDs we implement them in the cache hierarchy modifying the coherence protocol. 
For modeling the DRAM main memory we use DRAMSIM2~\cite{DRAMSim2}. We adapt the LLC read and write latencies according to the STT-RAM target. Both latencies and energy consumption values are obtained from NVSim~\cite{nvsim_12} for a 1MB (1 bank) cache and are illustrated in Table~\ref{tab:param}. For scaling the LLC to larger sizes, we multiply the leakage power by the number of cores.

Our experiments make use of the SPEC CPU2006 benchmark suite~\cite{SPEC}. When we evaluate our proposal in a single core scenario (LLC 1MB size) we employ \emph{reference} inputs and simulate 1 billion instructions from the checkpoint determined using PinPoints~\cite{Patil2004}. Note that results from 4 out of 29 benchmarks are not considered in the evaluation section due to experimental framework constraints. We also report results of 28 multiprogrammed mixes employing SPEC CPU2006 programs 
in 4, 8 and 16-CMP systems with 4, 8 and 16MB SLLC sizes, respectively. In all the cases, we fast forward 100M instructions, warm up caches for 200M instructions and then report results for at least 500M instructions per core.

For selecting the aforementioned multiprogrammed mixes, 
we employ the following methodology: we execute each benchmark alone, using an LLC of 1MB and without any reuse detector, and we measure the amount of LLC writes that it generates. We then obtain for each benchmark the \emph{number of writes to LLC per 1000 instructions} ratio (WPKI). Based on these values, we include each benchmark into the \emph{high}, \emph{medium} or \emph{low} category. Specifically, the \emph{high} category includes benchmarks with a WPKI higher than $8$, the \emph{medium} one those with a WPKI satisfying $1 < WPKI <8$ and finally, in the \emph{low} category we include the programs with a WPKI lower than $1$. Table~\ref{tab:writeperinst} shows this classification. Based on this classification, and as detailed in Section~\ref{sec:eval}, we build some mixes made up of programs with high WPKI, some with medium WPKI, some with low WPKI, and some combining applications from different WPKI categories trying to fill most of the combinations high-medium, high-low, medium-low and high-medium-low. Tables~\ref{tab:mix4C}, \ref{tab:mix8C} and \ref{tab:pmix16C} show the built mixes for the 4-core, 8-core and 16-core CMP systems, respectively, where for each mix the applications are sorted first by decreasing WPKI categories (from the high category to the low one) and then within each of the categories they are also sorted alphabetically. We employ the symbol ``;'' to separate applications from different categories and also the parenthesis to indicate how many instances of an application (when employ more than one) we are using in each mix.

\begin{table*}
	\begin{center}
		{\fontsize{7}{7}\selectfont
		\caption{SPEC 2006 multiprogrammed mixes for the 8-core CMP.\label{tab:mix8C}}
		\begin{tabular}{|c|c|}
			\hline
			\textbf{Mixes}&\textbf{Applications}\\
			\hline
				mix.H0 & bwaves, cactusADM, lbm, leslie3d, libquantum, mcf, milc, zeusmp \\
				\hline
				mix.H1 & bwaves, cactusADM(2), leslie3d(2), libquantum, mcf, milc\\
				\hline
				mix.H2 & bwaves, cactusADM, leslie3d, libquantum, mcf, milc(2), zeusmp\\
				\hline
				mix.H3 & cactusADM, lbm, leslie3d, libquantum(2), milc, zeusmp(2)\\
				\hline
				mix.M0 & astar, gcc, gobmk, hmmer, perlbench, soplex, wrf, xalancbmk\\
				\hline
				mix.M1 & bzip2, gcc, perlbench(2), soplex, wrf, xalancbmk(2), \\
				\hline
				mix.M2 & gcc, gobmk(2), hmmer, perlbench, soplex(2), xalancbmk\\
				\hline
				mix.M3 & astar, gobmk(2), hmmer, perlbench, soplex, wrf(2) \\
				\hline
				mix.L0 & calculix, GemsFDTD, gromacs, h264ref, namd, omnetpp, sphinx3, tonto \\
				\hline
				mix.L1 & GemsFDTD, gromacs, h264ref(2), namd, omnetpp(2), tonto \\
				\hline
				mix.L2 & gromacs(2), h264ref(2), namd, omnetpp, tonto(2)  \\
				\hline
				mix.L3 & calculix, GemsFDTD(2), gromacs, namd, omnetpp, tonto(2) \\
				\hline
				mix.HM0 & cactusADM, leslie3d, libquantum, mcf; bzip2, gobmk, soplex, xalancbmk \\
				\hline
				mix.HM1 & lbm, mcf, milc, zeusmp; gcc, hmmer, perlbench, wrf\\
				\hline
				mix.HM2 & bwaves, cactusADM, libquantum, milc; astar, gobmk, perlbench, soplex \\
				\hline
				mix.HM3 & cactusADM, lbm, leslie3d, zeusmp; gcc, gobmk, soplex, wrf\\
				\hline
				mix.HL0 & cactusADM, leslie3d, libquantum, mcf; calculix, GemsFDTD, namd, sphinx3 \\
				\hline
				mix.HL1 & lbm, mcf, milc, zeusmp; gromacs, h264ref, omnetpp, tonto \\
				\hline
				mix.HL2 & bwaves, lbm, libquantum, milc; GemsFDTD, h264ref, sphinx3, tonto\\
				\hline
				mix.HL3 & cactusADM, lbm, leslie3d, zeusmp; calculix, gromacs, namd, omnetpp\\
				\hline
				mix.ML0 & bzip2, gobmk, soplex, xalancbmk; calculix, GemsFDTD, namd, sphinx3 \\
				\hline
				mix.ML1 & astar, hmmer, perlbench, wrf; h264ref, gromacs, omnetpp, tonto \\
				\hline
				mix.ML2 & gcc, gobmk, perlbench, xalancbmk; GemsFDTD, gromacs, omnetpp, sphinx3\\
				\hline
				mix.ML3 & bzip2, hmmer, soplex, wrf; calculix, h264ref, namd, tonto \\
				\hline
				mix.HML0 & bwaves, cactusADM, leslie3d; gobmk, soplex, xalancbmk; GemsFDTD, sphinx3\\
				\hline
				mix.HML1 & lbm, libquantum, mcf; astar, perlbench; calculix, h264ref, namd \\
				\hline
				mix.HML2 & cactusADM, milc, zeusmp; hmmer, wrf, xalancbmk; h264ref, tonto\\
				\hline
				mix.HML3 & leslie3d, mcf, milc; perlbench, soplex; gromacs, namd, omnetpp\\
				\hline
		\end{tabular}}
	\end{center}
\end{table*}

\begin{table*}[htb]
	\begin{center}
{\fontsize{7}{8}\selectfont
		\caption{SPEC 2006 multiprogrammed mixes for the 16-core CMP.\label{tab:pmix16C}}{ 
		\begin{tabular}{|c|c|}
			\hline
			\textbf{Mixes}&\textbf{Applications}\\
			\hline
mix.H0 & bwaves(2), cactusADM(3), lbm, leslie3d(3), libquantum(2), mcf(2), milc(2), zeusmp\\
\hline
mix.H1 & bwaves(2), cactusADM, lbm(2), libquantum(2), mcf(3), milc(4), zeusmp(2)\\ 
\hline
mix.H2 & bwaves, cactusADM, lbm(3), leslie3d(4), libquantum(3), mcf(2), milc, zeusmp\\ 
\hline
mix.H3 & bwaves(3), cactusADM(2), lbm(2), leslie3d(2), libquantum, mcf, milc(2), zeusmp(3)\\ 
\hline
mix.HM0 & bwaves, cactusADM, lbm, leslie3d, libquantum, mcf, milc, zeusmp; astar, gcc, gobmk, hmmer, perlbench, soplex, wrf, xalancbmk\\
\hline
mix.HM1 & bwaves, cactusADM(2), leslie3d(2), libquantum, mcf, milc; bzip2, gcc, perlbench(2) soplex, wrf, xalancbmk(2)\\
\hline
mix.HM2 & bwaves, cactusADM, libquantum, leslie3d, mcf, milc(2), zeusmp; gcc, gobmk(2), hmmer, perlbench, soplex(2), xalancbmk\\
\hline
mix.HM3 & bwaves, cactusADM, lbm, leslie3d, libquantum, mcf, milc, zeusmp; astar, gobmk(2), hmmer, perlbench, soplex, wrf(2)\\
\hline
mix.M0 & astar(2), bzip2(2), gcc, gobmk(3), hmmer(2), perlbench(2), soplex, wrf(2), xalancbmk\\
\hline
mix.M1 & astar, bzip2(2), gcc, gobmk, hmmer, perlbench, soplex(2), wrf(3), xalancbmk(4)\\ 
\hline
mix.M2 & astar, bzip2, gcc, gobmk(3), hmmer(3), perlbench(2), soplex(4), xalancbmk\\
\hline
mix.M3 & astar(2), bzip2, gcc(3), perlbench, soplex(2), wrf(4), xalancbmk(3)\\
\hline
mix.L0 & calculix(2), GemsFDTD(2), gromacs(2), h264ref(2), namd(2), omnetpp(2), sphinx3(2), tonto(2)\\
\hline
mix.L1 & calculix, GemsFDTD(3), gromacs(3), h264ref, namd(2), omnetpp(2), sphinx3(3), tonto\\
\hline
mix.L2 & calculix(2), GemsFDTD(3), gromacs(3), h264ref(3), namd(3), tonto(2)\\ 
\hline
mix.L3 & calculix(4), gromacs(2), h264ref(2), omnetpp(4), sphinx3(2), tonto(2)\\ 
\hline
mix.HL0 & cactusADM, lbm, leslie3d, libquantum, mcf(2), milc, zeusmp; calculix, GemsFDTD, gromacs, h264ref, namd, omnetpp, sphinx3, tonto \\
\hline
mix.HL1 & bwaves, lbm(2), libquantum, mcf, milc(2), zeusmp; GemsFDTD, gromacs, h264ref(2), omnetpp, sphinx3, tonto(2) \\
\hline
mix.HL2 & cactusADM, lbm(2), leslie3d, mcf, milc, zeusmp(2); calculix, gromacs(2), h264ref, namd, omnetpp(2), tonto \\
\hline
mix.HL3 & bwaves, cactusADM, lbm, leslie3d, libquantum(2), mcf, milc; calculix, GemsFDTD(2), h264ref, namd, sphinx3(2), tonto \\ 
\hline
mix.ML0 & astar, bzip2, gobmk, hmmer, perlbench, soplex, wrf, xalancbmk; calculix, GemsFDTD, gromacs, h264ref, namd, omnetpp, sphinx3, tonto\\ 
\hline
mix.ML1 & astar(2), gcc, perlbench(2), soplex(2), xalancbmk(2); GemsFDTD(3), h264ref(2), tonto(2)\\ 
\hline
mix.ML2 & astar, bzip2, hmmer(2), perlbench, soplex, wrf(2); calculix, gromacs, h264ref(2), namd, omnetpp, tonto(2) \\
\hline
mix.ML3 & bzip2, gcc(2), gobmk(2), perlbench, soplex, xalancbmk(2); calculix, GemsFDTD, gromacs, namd, omnetpp, sphinx3(2) \\
\hline
mix.HML0 & bwaves, cactusADM, lbm, leslie3d, libquantum, mcf; astar, gobmk, perlbench, soplex, xalancbmk; calculix, GemsFDTD, h264ref, namd, sphinx3 \\
\hline
mix.HML1 & cactusADM, leslie3d, mcf, milc(2), zeusmp; hmmer, perlbench, soplex, wrf, xalancbmk; gromacs, h264ref, namd, omnetpp, tonto  \\
\hline
mix.HML2 & lbm, leslie3d, libquantum, mcf(2), milc; astar, perlbench(2), soplex; calculix, gromacs, h264ref, namd(2), omnetpp\\
\hline
mix.HML3 & bwaves, cactusADM(2), leslie3d, milc, zeusmp; gobmk, hmmer, soplex, wrf, xalancbmk(2); GemsFDTD, h264ref, sphinx3, tonto\\
\hline
		\end{tabular}}}
	\end{center}
\end{table*}

\noindent{\bf Energy model:} The DRAM energy is obtained directly from the simulator. For computing the LLC energy we employ a model that includes both dynamic and static contributions. The static component is calculated using NVSim~\cite{nvsim_12}, which reports the leakage number for 1MB LLC. Thus, we multiply that number by the execution time and the number of cores to obtain the total static energy. In the case of the dynamic component, we again use NVSim for determining the dynamic energy consumption per access to the LLC. Then, we compute the dynamic energy consumption as follows: 

 \small
\begin{align}
\label{eq:energy}
\emph{Dynamic Energy} = H_{LLC}*HE_{LLC}+W_{LLC}*WE_{LLC}+\nonumber\\
M_{LLC}*ME_{LLC}\nonumber\\
\end{align}
\normalsize
where H$_{LLC}$, W$_{LLC}$ and M$_{LLC}$ denote the number of hits, writes and misses in the LLC respectively, and HE$_{LLC}$, WE$_{LLC}$ and ME$_{LLC}$ correspond to the energy consumption of a hit, a write and a miss in the LLC respectively. 

\normalsize


\section{Evaluation}
\label{sec:eval}

This section compares how well RD and DASCA behave when managing an STT-RAM LLC, both in terms of performance and energy consumption of LLC and main memory. Single, four and eight-core systems are discussed in Sections~\ref{subsec:sc},~\ref{subsec:mc}, and~\ref{subsec:8_mc}, respectively. Finally, in Section~\ref{subsec:putting}, we show the results derived from the evaluation in a 16-core CMP and we analyze the performance of both approaches as the number of cores augments.

\begin{figure*}
\includegraphics{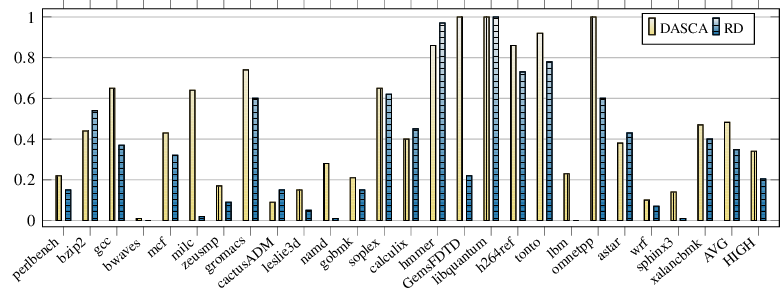}
\caption{Number of writes to the STT-RAM LLC normalized to the baseline: SPEC CPU2006 suite.}
\label{fig:writes_sc}
\end{figure*}
  
\subsection{Evaluation in a single-core scenario}
\label{subsec:sc}

First, we show the number of writes to the LLC that each evaluated proposal involves as well as the performance delivered. Then, we focus on the involved energy consumption in both the STT-RAM and the main memory according to the model detailed in Section \ref{sec:sim}. Finally, we discuss the obtained results. All the graphs shown in this section report individual data for each benchmark, adding at the right end the arithmetic mean considering all data (labeled as \emph{AVG}) or the geometric mean (labeled as \emph{GMEAN}) in the case of the performance metric, and also the arithmetic mean of the eight most write-intensive benchmarks according to Table~\ref{tab:writeperinst} or the corresponding geometric mean for the performance metric (labeled as \emph{HIGH} in both cases).

\subsubsection{Write filtering}
\label{subsubsec:wr_sc}

Figure~\ref{fig:writes_sc} illustrates the number of writes to the STT-RAM LLC generated by the DASCA scheme and our proposal (using a RD of 8K entries) normalized to a baseline system without any write reduction/filtering scheme.



As shown, our proposal significantly outperforms DASCA. Notably, in 20 out of 25 benchmarks evaluated the Reuse Detector exhibits higher ability in cutting the write traffic to the STT-RAM LLC. Overall, the block bypassing decisions commanded by RD reduce the number of LLC writes in the baseline system around 65\% whereas DASCA just achieves a 52\% reduction. In addition, if we zoom just in the 8 programs with highest WPKI numbers (those labeled as \emph{high} in Table~\ref{tab:writeperinst}), RD reduces the number of LLC writes by around 80\% with respect to the baseline, while DASCA cuts the write traffic by 66\%.

\subsubsection{Performance}
\label{subsubsec:perf_sc}

\begin{figure*}
\includegraphics{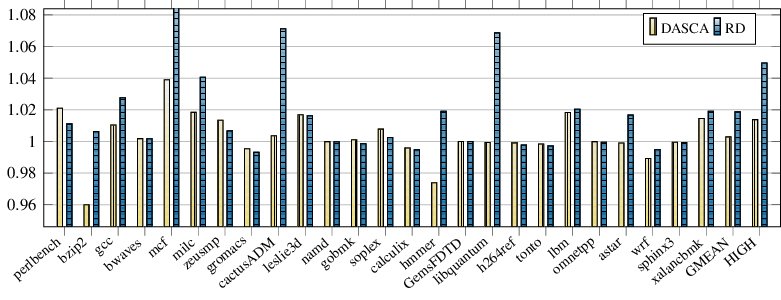}
\caption{Performance (Instructions per Cycle) normalized to the baseline: SPEC CPU2006 suite.}
\label{fig:perf_sc}
\end{figure*}

Apart from the goal of decreasing the STT-RAM LLC energy consumption (quantified later in this section), it is clear that energy efficiency should not come at the expense of a performance drop. Thus, to further evaluate the benefits of RD, Figure~\ref{fig:perf_sc} shows the performance (IPC) delivered.

Overall our scheme performs moderately better than DASCA: RD delivers 1.9\% performance improvement compared to the baseline while DASCA just improves IPC by 0.3\%.
If we focus on the write-intensive applications RD clearly outperforms DASCA, achieving performance improvements of 5\% and 1.4\%, respectively. This reveals, as we will confirm later in the multi-core environment, that our approach works especially well for those applications for which the number of writes performed to the LLC is high.

\subsubsection{Energy savings}
\label{subsubsec:energy_sc}

\begin{figure*}
\includegraphics{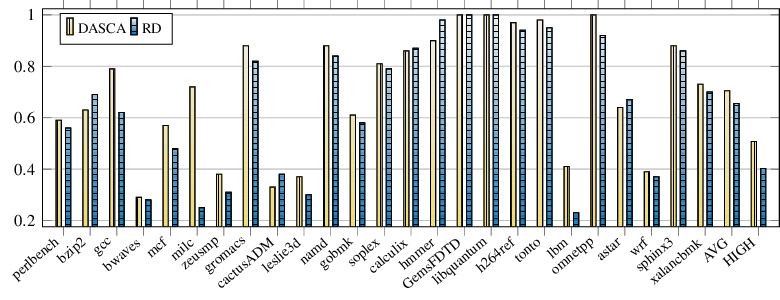}
\caption{Energy consumption in the STT-RAM LLC normalized to the baseline: SPEC CPU2006 suite.}
\label{fig:energy_sc}
\end{figure*}

\begin{figure}[h!]
\includegraphics{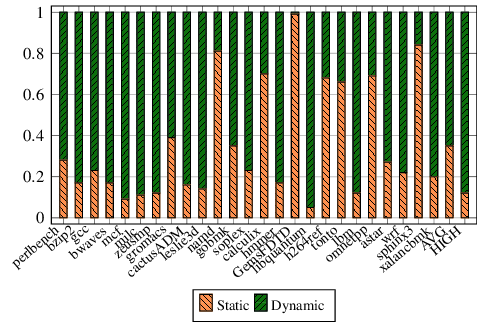}
\caption{Breakdown of energy consumption in the LLC into the static and dynamic contributions for the baseline in the single-core system.}
\label{fig:energysc_llc_break}
\end{figure}

Figure~\ref{fig:energy_sc} shows the total energy savings (adding both the dynamic and the static components) in the LLC. 
Overall, our proposal reports 34.5\% energy reduction compared to the baseline while DASCA reports 29.5\%. Considering only the write-intensive programs, the numbers are 60\% and 49\%, respectively. If we split the total energy savings with respect to the baseline into the dynamic and static parts, our proposal achieves 50\% of reduction in the dynamic part considering all the applications (68\% for the \emph{high} programs), while DASCA obtains 42\% (57\% for the \emph{high} benchmarks). As for the static part RD is able to obtain 2\% energy savings (around 5\% for the \emph{high programs}) while DASCA just achieves 0.3\% (1.4\% for the write-intensive applications). Note that avoiding LLC writes reduces dynamic energy, whereas increasing performance translates into static energy savings. It is also worth noting that, as Figure~\ref{fig:energysc_llc_break} illustrates, the dynamic energy consumption in the LLC of the baseline system is, for most of the applications evaluated, significantly higher than the static contribution.

Finally, we have also explored the impact on the energy consumption in the DRAM main memory. For the sake of simplicity, we do not show the results for all the applications. However, as expected, the DRAM energy reduction follows the trend of performance improvement. Overall, our proposal manages to reduce the DRAM energy consumption by 2\% (4.7\% for the write-intensive programs) with respect to the baseline while DASCA just improves the memory energy consumption by 0.2\% (1.1\% for the \emph{high} applications). 

\subsubsection{Discussion}
\label{subsubsec:disc_sc}

If we zoom into specific benchmarks, there are some special cases that deserve further detail to get a deeper insight. Note that globally, the relative trend shown in the amount of writes to the LLC between our approach and DASCA for each benchmark, is mainly held in the energy consumption differences, although modulated with the relative performance numbers. However, there are some few exceptions such as \emph{namd}, \emph{GemsFDTD} or \emph{omnetpp}, where RD is able to reduce the amount of LLC writes significantly more than DASCA but the energy savings and also the performance improvements obtained by both techniques are almost the same (and quite low compared to the baseline) in all these three cases. The reason is that these programs are three of the four benchmarks that exhibit the lowest values of WPKI, so although the write reductions that RD achieves in relative terms compared to DASCA is significant for these applications, the corresponding reduction in absolute values are very modest, and therefore the impact on the energy is almost negligible.

Also, in other applications such as \emph{mcf}, \emph{cactusADM} or \emph{hmmer}, our approach is able to report IPC numbers significantly higher than in DASCA, while both techniques exhibit quite similar write reduction capabilities. In order to explain that, first note that there are many different aspects involved in the system performance delivered. Among others, one key aspect is that reducing the amount of writes to the LLC is not sufficient in itself to guarantee performance improvements: although the main goals when bypassing blocks from the LLC to main memory 
are both to save energy and improve performance by increasing the hit rate in the LLC, obviously the bypassing may fail in the sense that a bypassed block could be referenced again soon, leading to a LLC miss and even a performance drop with respect to the case where bypassing is not carried out. Thus, for all these three benchmarks, the experimental data reveal that with our proposal the amount of hits in the LLC clearly outperforms both the baseline and the DASCA mechanism. Notably, the amount of LLC hits experienced in the \emph{cactusADM} and \emph{mcf} programs are 7.23x and 2x the values obtained in the baseline, while DASCA obtains 1.89x and 0.89x, respectively. Also, the amount of misses in the LLC is lower than that of the baseline and DASCA, with values ranging between 0.77-0.87x those obtained in the baseline. Considering all the evaluated benchmarks, RD is able to improve the amount of hits around 31\% with respect to the baseline (106\% if we only consider the write-intensive applications) while DASCA experiments only 5\% increment when considering all the benchmarks and 31\% for the \emph{high} applications. 

At a first glance, the behavior of the \emph{libquantum} application may seem somehow strange: Neither RD nor DASCA are able to significantly reduce the amount of writes to the LLC, but however this benchmark running under RD reports a performance improvement of 7\% with respect to the baseline while the performance remains largely unchanged under DASCA. In addition, and as one would expect since the number of bypasses is low, the number of hits in the LLC is practically the same in the three cases. The reason to explain the performance improvement lies in the LLC bank contention due to the write activity: this application is by far the most stalled one due to write contention. Thus, although the write reduction is very limited with our scheme, it is enough to reduce stalls with respect to the baseline by around 8\%, which in absolute numbers implies various millions of these kind of situations avoided, which leads to the performance improvement obtained.  

Conversely, although other benchmarks such as \emph{gromacs}, \emph{calculix} or \emph{wrf} exhibit moderate LLC writes reduction with RD and DASCA, they all perform worse than in the baseline. For these three programs the amount of hits experienced in the LLC is, in RD and DASCA, lower than in the baseline, which suggests that the bypassing performed is not efficient for these benchmarks. Recall that the energy savings achieved in the LLC as a consequence of the reduction in the number of writes performed in this cache level may be partially offset with the performance drop derived from the increment in the amount of LLC misses, as in these three programs occurs. Note also that, although the write operations are outside the critical path, the performance improvement derived from avoiding the long write operations may be mitigated if bank contention exists between the writes effectively performed.

\subsection{Evaluation in a 4-core CMP system}
\label{subsec:mc}

In this section we extend the previous single-core analysis to a more up-to-date environment: a multi-core scenario where the LLC is shared among different cores. 
For this purpose, we measure again the number of writes to the SLLC, the performance and the energy consumption in both the STT-RAM SLLC and the DRAM main memory for RD and DASCA and report results normalized to the baseline. However, due to the inherent non-determinism that all simulators exhibit (especially in multi-core environments, where the number of instructions executed across different schemes are not stable owing to the random interleaving among memory accesses of different programs) and for the sake of higher accuracy, we employ in this scenario, as well as in the 8 and 16-core CMP systems, the arithmetic mean of the number of writes and energy consumption (per application) but \emph{divided by the total number of instructions executed}. 
Note that, conversely, in the single-core scenario both kind of metrics match, since all the benchmarks execute the same amount of instructions (1B) in all the runs. 

We employ 28 mixes made up of applications from the SPEC CPU2006 suite chosen accordingly to the WPKI categories illustrated in Table~\ref{tab:writeperinst}. First, we randomly compose three groups of 4 mixes made up of applications belonging to just one WPKI category (mixes referred to as H$i$, M$i$ and L$i$ for high, medium and low WPKI respectively). Then, we build other 16 mixes merging applications with WPKI corresponding to different categories and trying to construct them in a balanced and homogeneous fashion. Again, the workload name encodes the WPKI categories of the applications. For example, HL2 is the third mix we build consisting of applications with high WPKI (2 in this case) and applications with low WPKI (other two). The detailed mixes are illustrated in Table~\ref{tab:mix4C}. Most graphs in this section report individual results for each mix, the arithmetic mean (AVG) considering all the mixes, just the 4 H$i$ mixes (HIGH), the 4 H$i$ and the 4 HM$i$ mixes together (H+HM), the 4 H$i$, the 4 HM$i$ and the 4 HML$i$ mixes together (H+HM+HML) and all the mixes including at least a high program (SomeH). Again, in the case of the performance metric we employ the geometric mean instead of the arithmetic one.

\subsubsection{Write filtering}
\label{subsubsec:wr_mc}

\begin{figure*}
\includegraphics{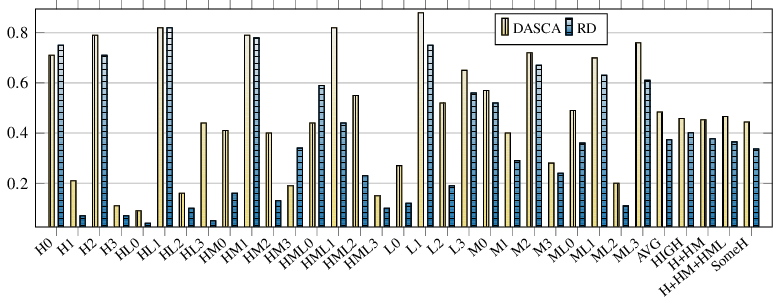}
\caption{Number of writes to the STT-RAM SLLC normalized to the baseline in the 4-core CMP system.}
\label{fig:writes_mc}
\end{figure*}

Figure~\ref{fig:writes_mc} illustrates the number of writes to the STT-RAM SLLC generated by using DASCA and an 8K-entry RD per core normalized to a baseline STT-RAM without any write reduction mechanism. 

The experimental results reveal that RD exhibits a significantly greater ability to decrease the amount of writes to the SLLC than DASCA. Notably, in 25 out of the 28 mixes evaluated RD outperforms DASCA. Overall, the number of writes in the baseline system gets reduced to 37\% by using RD, in contrast with DASCA which only achieves a 48\%. As for the write-intensive mixes the RD and DASCA makes around 40\% and 46\% of the writes the baseline performs, respectively.

\subsubsection{Performance}
\label{subsubsec:perf_mc}

\begin{figure*}
\includegraphics{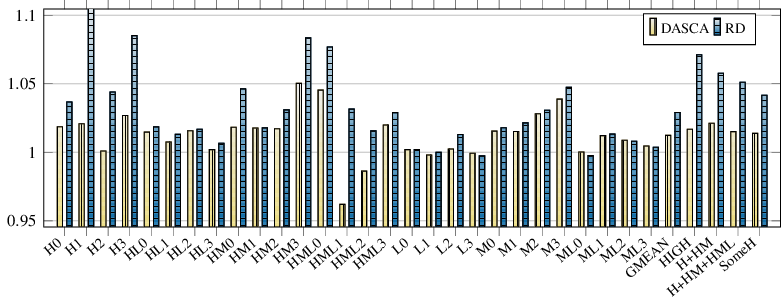}
\caption{Instruction throughput normalized to the baseline in the 4-core CMP system.}
\label{fig:perf_mc}
\end{figure*}

In order to evaluate performance when executing multiprogrammed workloads, we analyze the \emph{Instruction Throughput} (IT) and the \emph{Weighted Speedup} (WS) metrics. The IT metric is defined as the sum of all instructions committed per cycle in the entire chip ($\sum_{i=1}^{n} IPC_i$, being \emph{n} the number of threads), while the WS is defined as the slowdown experienced by each application in a mix, compared to its run under the same configuration when no other application is running on other cores ($\sum_{i=1}^{n} (IPC_{i}^{shared}/IPC_{i}^{alone})$). For the sake of simplicity and since in our context the WS does not constitute a metric as significant as IT, we do not show the WS results obtained. Anyway, these results follow an analogous trend to those obtained when we evaluate the instruction throughput.
Figure~\ref{fig:perf_mc} illustrates the IT that each evaluated policy delivers normalized to the baseline. 

As shown, RD moderately outperforms DASCA. This is a key contribution of RD, since our approach, managing to reduce the amount of writes to the SLLC to a greater extent than DASCA, is also able to deliver higher performance (which also allows to report higher energy savings in both the SLLC and the main memory as shown later). The data reveal that, overall, RD improves performance by around 3\% compared to the baseline, while DASCA just improves it by around 1.2\%. Moreover, we can observe that, in almost all of the 28 mixes evaluated (except mainly those mixes made up of benchmarks with a reduced WPKI, those labeled as \emph{low}, where the performance of both techniques essentially matches that of the baseline), our technique performs better. 
Zooming into particular mixes, the results reveal that RD performs especially better than DASCA in those mixes made up of write-intensive applications. Thus, our approach reports a performance improvement of more than 7\% when considering just the H$_i$ mixes while DASCA just reports 1.7\% IT improvement with respect to the baseline. Also, RD delivers significantly higher performance than DASCA and the baseline for those mixes which contain any application with high WPKI.

\subsubsection{Energy savings}
\label{subsubsec:energy_mc}

\begin{figure*}
\includegraphics{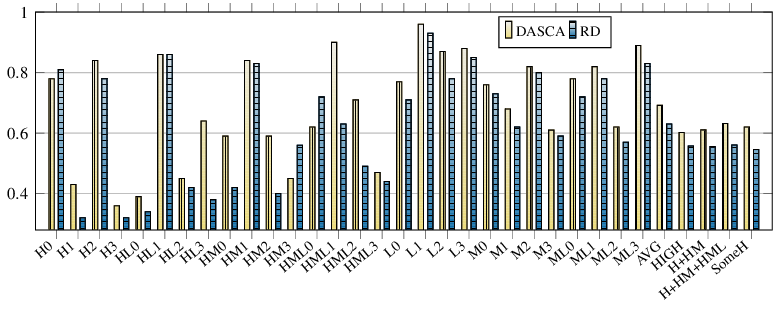}	
\caption{Energy consumption in the STT-RAM SLLC normalized to the baseline in the 4-core CMP system.}
\label{fig:energy_mc}
\end{figure*}

\begin{figure}
\includegraphics{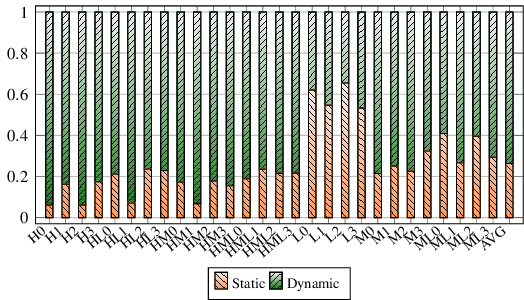}
\caption{Breakdown of energy consumption in the SLLC into the static and dynamic contributions for the baseline in the 4-core CMP system.}
\label{fig:energy_llc_break}
\end{figure}

Figure~\ref{fig:energy_mc} illustrates the energy savings in the SLLC. As in the single-core scenario, the graph follows a similar relative trend between our approach and DASCA to that observed in the write reduction numbers (Figure~\ref{fig:writes_mc}), just slightly modulated with the performance numbers since, as shown in Figure~\ref{fig:energy_llc_break}, the dynamic contribution to the energy consumption in the SLLC is higher than the static part (except in the mixes made up of applications with low WPKI only), so that the ability to reduce the amount of writes to the SLLC (dynamic contribution) impacts the total energy consumption more than the ability to improve performance, which mainly affects the static contribution. Overall, our proposal reports around 37\% energy reduction in the STT-RAM SLLC compared to the baseline while DASCA reduces it by around 31\%. If we zoom into the write-intensive mixes, both RD and DASCA are able to save around 45\% and 39\% of SLLC energy consumption, respectively. If we break the SLLC energy numbers down into the static and dynamic contributions, 
our results reveal that, overall, RD is able to reduce --considering all mixes-- the static energy part by around 2.7\% with respect to the baseline (around 6\% for the write-intensive mixes) while DASCA reduces the static contribution by 1.2\% (1.7\% for the \emph{high} mixes). In addition, our approach reports dynamic energy savings of around 50\% (51\% for the \emph{high} mixes) while DASCA numbers are 42\% (46\% for the \emph{high} mixes).

\begin{figure*}
\includegraphics{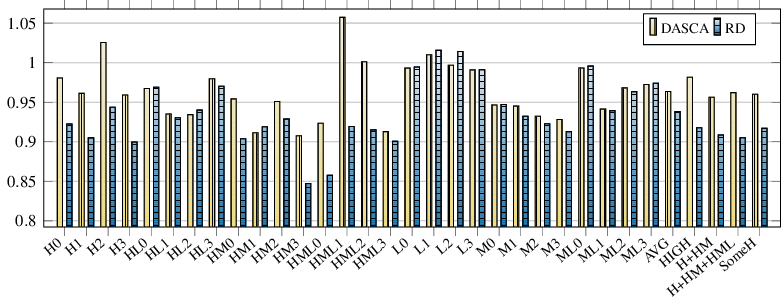}	
\caption{Energy consumption in the DRAM normalized to the baseline in the 4-core CMP system.}
\label{fig:energy_dram_mc}
\end{figure*}

Also, we explore the energy savings obtained in the DRAM main memory, where the leakage contribution has far greater significance than in the STT-RAM SLLC, so that the trends obtained essentially follow those of the IT graph, but inverted (higher performance translates into lower DRAM energy consumption). Figure~\ref{fig:energy_dram_mc} illustrates that RD manages to additionally reduce the energy consumption of the main memory by around 6.2\% on average compared to the baseline (8.3\% for the write-intensive mixes), while DASCA barely reaches a 3.6\% energy reduction (around 2\% for the \emph{high} mixes), mainly due to the higher performance improvement that our proposal exhibits. 


\subsubsection{Discussion}
\label{subsubsec:disc_mc}

\begin{figure*}
\includegraphics{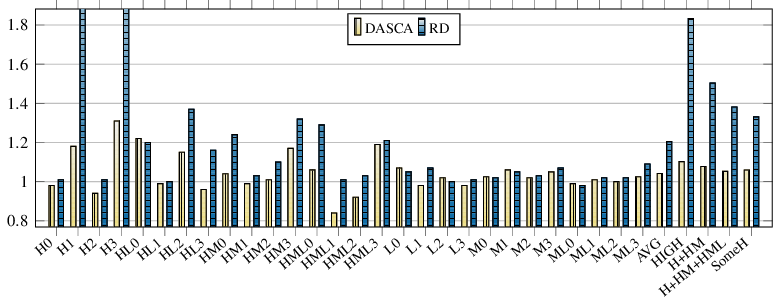}
\caption{Number of STT-RAM SLLC hits per kilo instruction normalized to the baseline in the 4-core CMP system.}
\label{fig:hits_mc}
\end{figure*}

For the sake of clarity, we next explain where the performance improvements of our technique come from. First, as Figure~\ref{fig:writes_mc} illustrated earlier, the write reductions to the SLLC that RD achieves are greater than those of DASCA. Second, and more importantly, as Figure~\ref{fig:perf_mc} reveals, the bypasses dictated by RD translate into more performance than that of DASCA. As in the single-core scenario, the rationale behind that is related with the hit rate experimented in the SLLC with both schemes. Figure~\ref{fig:hits_mc} illustrates the number of hits in the SLLC per kilo instruction that each mix experiments normalized to the baseline. 

The results reveal that in most of the mixes evaluated the amount of hits in the SLLC is higher under our approach than using DASCA. Again, this is especially evident for the case of the mixes including write-intensive applications such as H1, H3 and HL2 where the number of hits is 2.87x, 2.45x and 1.37x those of the baseline, respectively. This is the key to explain our performance improvements: the efficient management of the SLLC contents by exploiting the reuse locality. In addition, there are other factors that also contribute to the throughput gain such as less write operations to the SLLC, less main memory accesses, and increased row buffer hit rates. In order to perform a deeper comparison between RD and DASCA, Table~\ref{tab:comp} recaps the average values of different metrics involved in the performance delivered by RD and DASCA, normalized to those of the baseline. As shown, our scheme improves DASCA and the baseline (especially in the data from write-intensive mixes) in all the metrics considered.

\begin{table*}
	\begin{center}
		\tbl{Average values of different metrics normalized to the baseline in the 4-core CMP system.\label{tab:comp}}{ 
		\resizebox{0.8\textwidth}{!}{%
		\begin{tabular}{|l|C{3.5em}|C{3.5em}|C{7.5em}|C{3.5em}|C{3.5em}|C{8.5em}|}
			\hline
			 \backslashbox{\textbf{Policies}}{\textbf{Metrics}}&\textbf{SLLC Misses}&\textbf{SLLC Hits}&\textbf{Row buffer Read Hit Rate}&\textbf{DRAM reads}&\textbf{DRAM Writes}&\textbf{Bank contention in SLLC}\\
				\hline
				\textbf{DASCA (All/High)} & 1.01/1.05 & 1.04/1.10 & 1.03/1.00 & 1.01/1.05 & 1.04/1.06 & 0.45/0.16\\
				\hline
				\textbf{RD (All/High)} & 0.94/0.95 & 1.20/1.83 & 1.05/1.01 & 0.98/0.98 & 0.96/0.94 & 0.29/0.08\\ 
				\hline
		\end{tabular}}}
	\end{center}
\end{table*}

As in the single-core scenario, next we zoom into particular mixes that need further detail to get a better understanding. First, in some mixes such as H0, HM3 or HML0 we can observe that the DASCA scheme is able to reduce the amount of writes to the SLLC and also the energy consumption in the STT-RAM more than our scheme does (Figures~\ref{fig:writes_mc} and \ref{fig:energy_mc}). Conversely, the RD manages to deliver more throughput than DASCA (Figure~\ref{fig:perf_mc}). However, these performance improvements our approach achieves are not enough to offset the higher energy savings in the SLLC that the DASCA scheme reports for these mixes as a consequence of the lower number of writes to the STT-RAM. 

Second, data for mix L2 reveal that the RD is able to reduce the amount of writes to the SLLC much more than DASCA with respect to the baseline (81\% vs. 48\%). However, this great difference translates into just 22\% of energy savings in RD vs. 13\% of DASCA. As shown, the difference between both policies has been significantly reduced due to the low contribution of the dynamic energy to the total energy consumption in the SLLC that this mix exhibits, as Figure~\ref{fig:energy_llc_break} illustrates.

\subsubsection{Sensitivity to Reuse Detector size}
\label{subsubsec:sense}

\begin{figure*}
\includegraphics{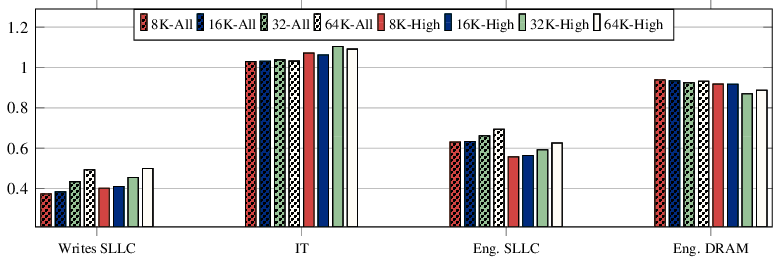}	
\caption{Writes to SLLC, IT and energy consumption in both SLLC and main memory normalized to the baseline for different RD sizes per core in the 4-core CMP system.}
\label{fig:sense_sc}
\end{figure*}

The RD size is a key design aspect of our proposal.  
In order to evaluate its impact we show in Figure~\ref{fig:sense_sc} the amount of writes to the SLLC, the Instruction Throughput, 
and the energy consumption in both the SLLC and the main memory for different RD sizes per core, namely 8K, 16K, 32K and 64K entries. 

As shown, the major impact is observed on the capability to reduce the number of writes in the SLLC, ranging from an average reduction of 63\% with respect to the baseline when an 8K-entry RD per core is employed (60\% for the write-intensive mixes) to a reduction of around 51\% for a 64K-entry RD per core (50\% for the \emph{high mixes}). Note that maybe these data might appear contradictory at first sight. However, they are not: As the size of RD increases, it also augments the probability that a block finds its tag in the RD, so the probability of bypassing decreases, leading to minor reduction of writes to the SLLC. We can also observe a moderate impact on the average energy consumed in the SLLC, with values in the range 63-69\% as the size of RD gets increased: again, note that these numbers follow a similar trend to that exhibited by the amount of writes. Finally, the impact over the performance and the energy consumption of the main memory is much reduced, falling the average IT variations into a small range of 1\% (4\% for the write-intensive mixes) and the average DRAM energy variations into a range of 1.5\% (5\% for the write-intensive mixes). 

\subsubsection{Overhead analysis}
\label{subsubsec:overhead}

In Section~\ref{subsubsec:details} we outlined that an 8K-entry RD for a 1MB LLC requires an extra storage of 14KB, which represents a 1.37\% overhead with respect to the LLC size. In this section we previously noted that for the 4-CMP system under evaluation (4MB SLLC) we employ an 8K-entry RD per core. The reason is that we are maintaining for each evaluated system the 1.37\% overhead with respect the SLLC size. 
Therefore, in the 8-CMP evaluated later, we also employ an 8K-entry RD per core. Hence, the total extra storage (overhead) of RD is 56KB and 112KB for the 4-CMP and 8-CMP systems respectively, representing in all the cases a 1.37\% overhead with respect to the SLLC size.    

\subsubsection{RD in a two-level cache hierarchy}
\label{subsubsec:2_levels}

We have evaluated the operation of our proposal in a three-level cache hierarchy since most current processors employ this configuration. Furthermore, two private levels are more likely to filter better the temporal locality than using just one private level. However, for a fair comparison, we have also evaluated our proposal and the DASCA scheme in a configuration with just two cache levels. Notably, we reproduce the same configuration (4-CMP) used by the authors in~\cite{dasca_14} when presenting the DASCA technique (32 KB IL1 and DL1 as private caches and a 1MB per core shared L2 cache). Table~\ref{tab:2levels} illustrates the main results.

\begin{table*}
	\begin{center}
		\tbl{Average values of different metrics normalized to the baseline\label{tab:2levels} in a 4-core CMP system with two cache levels.}{ 
		\resizebox{0.8\textwidth}{!}{%
		\begin{tabular}{|l|C{3.5em}|C{7em}|C{8em}|C{8em}|C{3.5em}|}
			\hline
			 \backslashbox{\textbf{Policies}}{\textbf{Metrics}}&\textbf{Writes SLLC}&\textbf{Instr. Throughput}&\textbf{Energy consumpt. SLLC}&\textbf{Energy consumpt. DRAM}&\textbf{SLLC Hits}\\
				\hline
				\textbf{DASCA (All/High)} & 0.70/0.79 & 1.01/1.03 & 0.81/0.86 & 0.97/0.96 & 0.98/1.00\\
				\hline
				\textbf{RD (All/High)} & 0.59/0.76 & 1.03/1.04 & 0.75/0.83 & 0.95/0.95 & 0.98/1.00\\ 
				\hline
		\end{tabular}}}
	\end{center}
\end{table*}

As shown, RD maintains higher capability than DASCA (around 11\% higher) in reducing the amount of writes to the SLLC. However, as expected, the amount of writes avoided (and also the hits experienced in the SLLC) is significantly lower than that exhibited in an scenario with 3 cache levels. Recall that this is due to the fact that with two cache levels only, most temporal locality has not been filtered, so that the reuse locality can not be fully exploitable. Also, as a consequence of this lower capability in cutting the write traffic to the SLLC, the energy savings achieved in the shared L2 are significantly lower than those obtained with three cache levels, although RD still reports better numbers than DASCA. Finally, RD again improves the Instruction Throughput to a greater extent than DASCA, and consequently also delivers higher energy savings in the main memory. Note that we have also evaluated 28 mixes in this configuration following the same criteria explained earlier, but they are not exactly the same as in the three-level cache hierarchy experiments since the WPKI values that the benchmarks exhibit do not match those of the three-level configuration and therefore some programs changed the category (high, medium or low) in which they were classified. 


\subsection{Evaluation in an 8-core CMP system}
\label{subsec:8_mc}

In this section we illustrate and analyze the main results obtained when using RD and DASCA in an 8-core CMP system with an 8MB SLLC. Like in the previous section, in this scenario we create 28 mixes following the same criteria as in a 4-CMP system. The mixes evaluated are shown in Table~\ref{tab:mix8C}. 
Given that a detailed analysis of the 8-core system would show similar results as the 4-core scenario, in this section we will not zoom into details but will only describe the average results and main trends.

\subsubsection{Write filtering}
\label{subsubsec:wr_8mc}

\begin{figure*}[htb]
\includegraphics{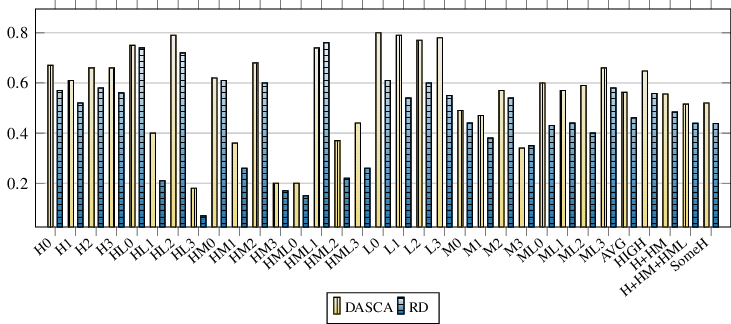}
\caption{Number of writes to the STT-RAM SLLC normalized to the baseline in the 8-core CMP system.}
\label{fig:writes_8mc}
\end{figure*}

Figure~\ref{fig:writes_8mc} illustrates the number of writes to the STT-RAM SLLC generated with DASCA and with RD (assuming an 8K-entry RD per core). Both schemes are normalized to a baseline STT-RAM without any content selection mechanism. 

Similarly to the results for the 4-core scenario, the experimental results reveal that RD just performs 46\% of the writes made in the baseline scheme, whereas DASCA produces 56\% of the writes that the baseline did. For the write-intensive mixes, RD and DASCA reduce the amount of writes compared to the baseline in 44\% and 35\% respectively.

\subsubsection{Performance}
\label{subsubsec:perf_8mc}

\begin{figure*}[htb]
\includegraphics{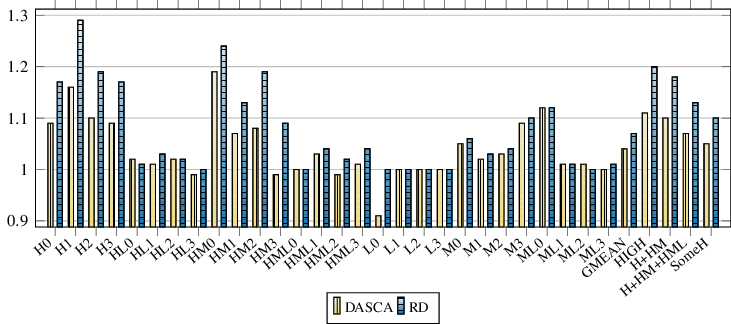}
\caption{Instruction throughput normalized to the baseline in the 8-core CMP system.}
\label{fig:perf_8mc}
\end{figure*}

As we did in Section~\ref{subsubsec:perf_mc}, we employ the \emph{Instruction Throughput} (IT) to evaluate the performance when executing multiprogrammed workloads. Figure~\ref{fig:perf_8mc} illustrates the IT that each evaluated policy delivers normalized to the baseline. 

Similarly to the results obtained for a 4-core CMP system, RD outperforms DASCA in the 8-core scenario. Moreover, in the 8-core scenario, higher performance improvements are achieved in both schemes over the baseline. The results reveal that RD improves performance by around 7\% compared to the baseline, while DASCA improves it by around 4\%. As for write-intensive mixes, RD improves the baseline by 20\% and DASCA by 11\%. As shown in Figure~\ref{fig:perf_8mc}, RD significantly overcomes DASCA and the baseline scheme in those mixes which contain any application with high WPKI.

\subsubsection{Energy savings}
\label{subsubsec:energy_8mc}

\begin{figure*}[htb]
\includegraphics{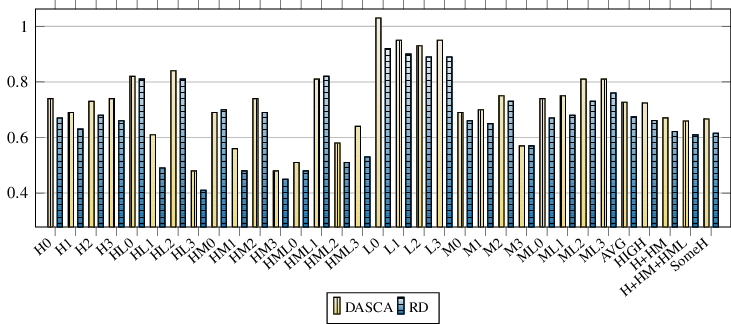}
\caption{Energy consumption in the STT-RAM SLLC normalized to the baseline in the 8-core CMP system.}
\label{fig:energy_8mc}
\end{figure*}

\begin{figure}[htb]
\includegraphics{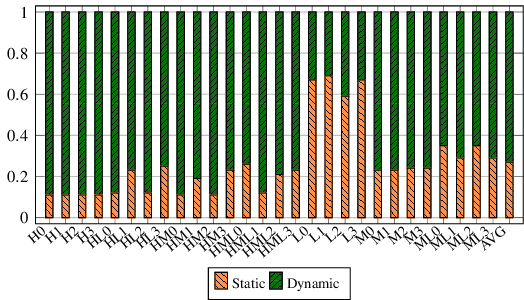}
\caption{Breakdown of energy consumption in the SLLC into the static and dynamic contributions for the baseline in the 8-core CMP system.}
\label{fig:energy_llc_break8c}
\end{figure}

Figure~\ref{fig:energy_8mc} illustrates the energy savings in the shared LLC. In general, the results in the 8-core scenario follow the trend observed for the 4-core environment. Specifically, RD reports around 32.5\% energy reduction in the STT-RAM SLLC compared to the baseline while DASCA reduces energy by around 27\%. In the case of write-intensive mixes, both RD and DASCA reduce the SLLC energy consumption by 34\% and 27.5\%, respectively. Analyzing  the static and dynamic contributions on the SLLC energy consumption, 
overall, RD is able to reduce --for all mixes-- the static energy part by around 6\% with respect to the baseline (around 15\% for the write-intensive mixes) while DASCA reduces the static contribution by 3.6\% (9.5\% for the \emph{high} mixes). In addition, our approach reports dynamic energy savings of around 43\% (36\% for the \emph{high} mixes) while DASCA numbers are 36\% (30\% for the \emph{high} mixes).
Note that mixes made up of applications with low WPKI exhibit the lowest energy savings across the board. This is consistent with the modest write reduction they report and especially with the high contribution of the static part to the total SLLC energy consumption that they exhibit, as Figure~\ref{fig:energy_llc_break8c} shows.

\begin{figure*}[htb]
\includegraphics{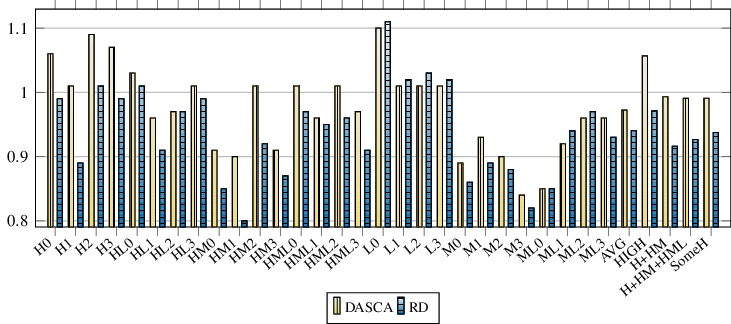}
\caption{Energy consumption in the DRAM normalized to the baseline in the 8-core CMP system.}
\label{fig:energy_dram_8mc}
\end{figure*}

Figure~\ref{fig:energy_dram_8mc} illustrates the energy savings obtained in the DRAM main memory, where it is shown that RD reduces the energy consumption of the main memory by around 6\% on average compared to the baseline (3\% for the write-intensive mixes), while DASCA reaches a 2.8\% energy reduction and actually wastes more energy, around 6\%, for the \emph{high} mixes. This energy waste may look surprising, given that DASCA is able to reduce the number of writes with respect to the baseline by 35\% and to deliver a performance improvement higher than 10\%. However, this can be explained by the fact that DASCA suffers a very significant increase in the amount of SLLC misses, which translates into high values of DRAM accesses (as shown in Table~\ref{tab:comp8c}). 

\subsubsection{Discussion}
\label{subsubsec:disc_8mc}

As in the 4-core configuration, in this section we explain the reasons for the higher performance improvement achieved in our technique (RD) against DASCA in the 8-core scenario.

\begin{figure*}[h!]
\includegraphics{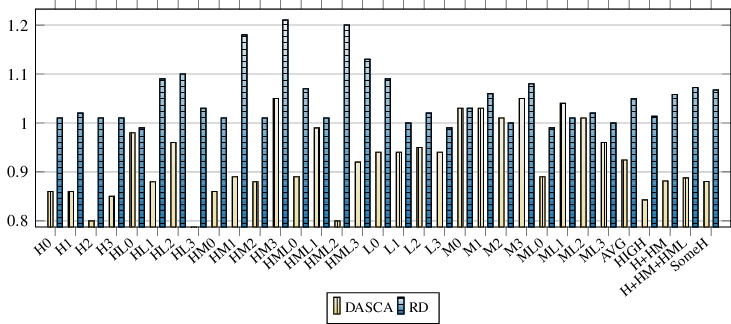}
\caption{Number of STT-RAM SLLC hits per kilo instruction normalized to the baseline in the 8-core CMP system.}
\label{fig:hits_8mc}
\end{figure*}

As we already reasoned in the previous section, the better performance of RD is due to several factors, being the most important one the high efficiency achieved from the reuse locality exploitation. For demonstrating that fact, Figure~\ref{fig:hits_8mc} shows the number of hits in the SLLC per kilo instruction that each mix experiments normalized to the baseline. As the figure shows, our approach achieves in most mixes a higher or much higher number of hits than DASCA, which confirms that RD uses a more efficient policy than DASCA.


In addition to the hit rate improvement, there are other metrics that also justify achieving a better performance, such as SLLC misses, DRAM reads and writes, row buffer read hit rate and bank contention in the SLLC. All these metrics are shown in Table~\ref{tab:comp8c}, for both RD and DASCA and also for both all and write-intensive mixes. Note that the RD beats DASCA in all the metrics considered. 
\begin{table*}
	\begin{center}
		\tbl{Average values of different metrics normalized to the baseline in the 8-core CMP system.\label{tab:comp8c}}{ 
		\resizebox{0.8\textwidth}{!}{%
		\begin{tabular}{|l|C{3.5em}|C{3.5em}|C{7.5em}|C{3.5em}|C{3.5em}|C{8.5em}|}
			\hline
			 \backslashbox{\textbf{Policies}}{\textbf{Metrics}}&\textbf{SLLC Misses}&\textbf{SLLC Hits}&\textbf{Row buffer Read Hit Rate}&\textbf{DRAM reads}&\textbf{DRAM Writes}&\textbf{Bank contention in SLLC}\\
				\hline
				\textbf{DASCA (All/High)} & 1.08/1.30 & 0.92/0.84 & 1.00/0.99 & 1.08/1.30 & 1.09/1.21 & 0.40/0.13 \\
				\hline
				\textbf{RD (All/High)} & 0.98/1.00 & 1.05/1.01 & 1.02/1.04 & 1.00/1.06 & 1.02/1.05 & 0.24/0.07\\ 
				\hline
		\end{tabular}}}
	\end{center}
\end{table*}

\subsubsection{Sensitivity to Reuse Detector size}
\label{subsubsec:sense8c}

\begin{figure*}[t!]
\includegraphics{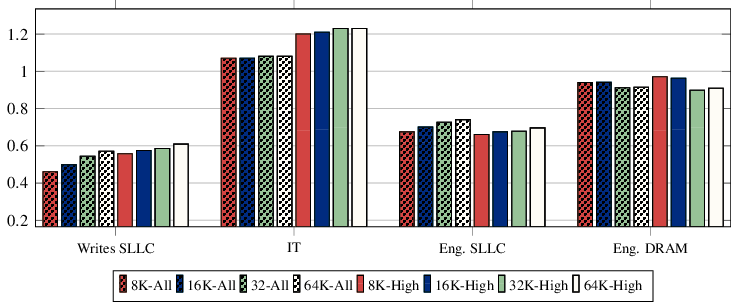}	
\caption{Writes to SLLC, IT and energy consumption in both SLLC and main memory normalized to the baseline for different RD sizes per core in the 8-core CMP system.}
\label{fig:sense_sc8c}
\end{figure*}

Given that the RD size is a determining factor in our proposal, and as done in the 4-CMP system, in Figure~\ref{fig:sense_sc8c} we show the amount of writes to the SLLC, the Instruction Throughput, and the energy consumption in both the SLLC and the main memory for different RD sizes per core, namely 8K, 16K, 32K and 64K entries. 

The trends are very similar to those observed in the 4-core scenario. Notably, a significant impact is observed on the capability to reduce the number of writes in the SLLC, especially for the All mixes, whereas a moderate (or even negligible in some cases) impact is seen on the average energy consumed in that cache level or main memory and performance of the overall system.

\subsection{RD performance in multi-core scenarios}
\label{subsec:putting}

In this section we briefly inspect the main hints about the performance of RD when we augment the number of cores. So far, we have evaluated the RD in systems with one, four and eight cores. In order to further explore the RD behavior we have also evaluated our proposal in a system with 16 cores, employing the mixes shown in Table~\ref{tab:pmix16C}.

In Table~\ref{tab:cores} we recap the main numbers derived from RD and DASCA evaluation across the different configurations (for the sake of simplicity, we show just the numbers considering all applications or mixes, not those corresponding to the write-intensive ones). Notably, we illustrate the average LLC write reduction capability, energy savings in the LLC, the performance delivered and the energy savings in the DRAM.

\begin{table*}
	\begin{center}
		\scriptsize
		\tbl{Average values of different metrics normalized to the baseline across different configurations.\label{tab:cores}}{ 
		\resizebox{0.8\textwidth}{!}{%
		\begin{tabular}{|l|C{7.5em}|C{7.5em}|C{8em}|C{7.5em}|}
			\hline
			 \backslashbox{\textbf{Scenario}}{\textbf{Metrics}}&\textbf{LLC Write reduction (\%) (DASCA/RD)}&\textbf{LLC Energy Savings (\%) (DASCA/RD)}&\textbf{Performance improvement (\%) (DASCA/RD)}&\textbf{DRAM Energy Savings (\%) (DASCA/RD)}\\
				\hline
				\textbf{Single core} & 51.8 / 65.2 & 29.5 / 34.5 & 0.3 / 1.9 & 0.2 / 2.0 \\
				\hline
				\textbf{4-core CMP} & 51.6 / 62.7 & 30.8 / 37.0 & 1.2 / 2.9 & 3.6 / 6.2 \\
				\hline
				\textbf{8-core CMP} & 43.8 / 54.0 & 27.3 / 32.5 & 3.7 / 6.7 & 2.8 / 6.0 \\
				\hline
				\textbf{16-core CMP} & 38.6 / 46.7 & 26.0 / 30.5 & 10.2 / 14.5 & 4.0 / 7.9 \\
				\hline
		\end{tabular}}}
	\end{center}
\end{table*}

As shown, the write reduction capability in percentage terms gets reduced with the number of cores for both RD and DASCA. 
However, despite this decrease in the write filtering numbers, the most important consequence derived from this aspect, i.e. the net energy savings in the LLC, essentially remains in the range of 30-37\% for RD, slightly decreasing with the number of cores as a consequence of the increment in the number of writes, but keeping the benefit with respect to DASCA largely unchanged. Moreover, the performance improvement increases as the number of cores augments for both RD and DASCA, reporting significant numbers especially for the 8-core and 16-core system. Indeed, the performance improvement that RD reports with respect to the DASCA scheme also increases with the number of cores (which is especially significant in the case of write-intensive mixes for the 16-CMP system, where RD is able to increase the performance of the baseline by around 38\%, whereas DASCA only achieves 24\% of improvement). This is due to the fact that the difference in LLC misses reported by both approaches increases as the number of cores augments (notably, these numbers are 1\%, 6\%, 10\% and 13\% for 1, 4, 8 and 16-core systems respectively, always reporting RD the best value) and also the difference in other factors as the number of hits in the LLC, the number of DRAM reads and writes or the row buffer hit rate also increase, benefiting RD more with the number of cores. Finally, and derived from the higher performance delivered, the same trend is observed in the energy savings experienced in the main memory (where it is worthy to note that, in the 16-core scenario, RD manages to reduce the energy consumption of the baseline by more than 15\% for the write-intensive mixes, while DASCA hardly achieves 8\%). Overall we can conclude that, as the number of cores increases, RD is able to significantly increase the performance delivered (and also to increase the energy savings achieved in the main memory) at the expense of a slight decrease in the LLC energy savings obtained.

\section{Related Work}
\label{sec:related}

To address the problems of energy consumption and performance of STT-RAM SLLCs, in the last years different researchers have proposed solutions aiming to reduce either the amount of writes or the per-write energy.



A great body of work mainly tries to cut the write traffic to the STT-RAM: 
In~\cite{wang2013oap} the authors propose an obstruction-aware cache management policy called OAP. OAP monitors the cache to periodically detect LLC-obstruction processes, and manage the cache accesses from different processes, so that when an LLC-obstruction is detected the data is forwarded to the next cache level or Main Memory as appropriate. In~\cite{Rasquinha2011} two techniques are proposed to reduce the number of writes to a last level (L2) STT-RAM cache and also save energy. The first one adds a small cache between L1 and L2 --called write-cache (WC)-- which is mutually exclusive with L2 and stores only the dirty lines evicted from L1. On a cache access, both L2 and WC are accessed in parallel. The write misses are allocated in WC and the load misses are allocated in L2. WC reduces the number of L2 writes by absorbing most of the L1 writebacks. Other authors propose a coding scheme for STT-RAM last level cache based on the concept of value locality. They reduce switching probability in cache by swapping common patterns with limited weight codes to make writes less often as well as more uniform~\cite{yazdanshenas2013coding}.  
Other techniques~\cite{jung2013energy} rely on the observation that on average, a large fraction of bytes and words written to the L2 cache are only zero-valued data. Based on this, this technique adds additional ``all-zero-data'' flags in the tag arrays at the granularity of a single byte and a single word. Before any cache write, the data value is checked. If the all-zero bytes or words are detected, the corresponding flags are set and only the non-zero bytes or words are written. During a cache read operation, only the non-zero bytes or words are read and then the actual data are constructed by combining the information from the all-zero flags. 
Another proposal~\cite{park2012future} logically divides the STT-RAM cache line into multiple partial lines. In L1 cache, a history bit is kept for each partial line to track which partial lines have changed. Using this information, when a dirty L1 block is written to last level cache, only those partial lines which have been changed are written. Other authors propose techniques for mitigating the write pressure caused due to prefetching in STT-RAM based LLC~\cite{mao2013coordinating}. One of these techniques prioritizes different types of LLC requests such as load, store, prefetch, or write back, etc. based on their criticality. The critical requests are assigned a high priority and hence, they are served earlier. In multicore systems, the excessive requests generated from a cache-intensive program may block those generated from a cache-unintensive program which may lead to its starvation. To address this, they propose another technique which prioritizes the requests from a cache-unintensive program, so that they are served promptly. Also, authors in~\cite{changimpact} analyze the cache coherence protocols impact in the number of write to a LLC based on STT-RAM, showing that the protocols with a owned state (MOESI and MOSI) reduce the number of writes to LLC.

Another body of work mainly deals with performance of STT-RAM caches: In~\cite{jog2012cache} a cache revive technique to calculate retention time is proposed. Some cache blocks retain data even after completion of retention time. The retention time is chosen so that it will minimize the number of unrefreshed cache blocks. Other authors propose the use of STT-RAM to design combinational logic, register files and on-chip storage (I/D L1 caches, TLBs and L2 cache)~\cite{guo2010resistive}. Also, to hide the write latency of STT-RAM, they propose subbank buffering which allows the writes to complete locally within each sub-bank, while the reads from other locations within the array can complete unobstructed. They show that by carefully designing the pipeline, the STT-RAM based design can significantly reduce the leakage power, while also maintaining the performance level close to the CMOS design. Also, an STT-RAM cache design for lower level caches where different cache ways are designed with different retention periods is proposed in~\cite{sun2011multi}. For example, in a 16-way cache, way 0 is designed with a fast STT-RAM design with low retention period and the remaining 15 ways are designed with a slow STT-RAM design which has higher retention period. Their technique uses hardware to detect whether a block is read or write intensive. The write intensive blocks are primarily allocated to way 0, while the read intensive blocks are allocated to the other ways. Also, to avoid refreshing dying blocks in way 0, their technique uses data migration to move such blocks to banks with higher retention period. Finally, a write-buffer design to address the long write latency of last level (L2) STT-RAM cache is proposed in~\cite{sun2009novel}. The L2 may receive a request from both L1 and the write buffer. Since read latency of STT-RAM is smaller than the write latency and also reads are performance-critical, the buffer uses a read-preemptive management policy, which ensures that a read request receives higher priority than a write request. The authors also propose a hybrid SRAM and STT-RAM cache design which aims to move the most write-intensive blocks to SRAM.

It is worth mentioning that we compare our RD scheme just against DASCA due to three reasons: 1) DASCA, which is a more recent proposal than all the aforementioned works, is the closest work to ours in the sense that both schemes try to reduce the energy consumption of an STT-RAM LLC by bypassing write operations predicted to be dead (or not showing reuse), 2) some of the mentioned approaches are already evaluated in the DASCA paper, such as~\cite{wang2013oap}, being clearly outperformed by the DASCA mechanism, and 3) some other schemes, such as ~\cite{yazdanshenas2013coding,jung2013energy,park2012future,mao2013coordinating,jog2012cache,sun2011multi,sun2009novel}, although addressing the same problem, are completely orthogonal to our approach. Thus, the RD could be built on top of them, making a direct comparison meaningless. As an example, in ~\cite{yazdanshenas2013coding} the authors address the STT-RAM write energy problem at a circuit level, trying to reduce the number of writes and also to balance the wear among the cells. The proposal relies in identifying the most frequent values stored in the LLC (value locality) and encoding these patterns to reduce the number of writes. Thus, unlike our proposal, this approach operates at a bit-level. 

\section{Conclusions}

In this paper we have addressed the main constraints of conventional SRAM last-level caches: power-hungry operation and inefficient management. In order to overcome these drawbacks we propose to employ a STT-RAM SLLC where its contents are selected according to a \emph{Reuse Detector} which exploits the reuse locality of the stream of references arriving at the SLLC. The Reuse Detector is a hardware component that tracks block reuse and determines, according to its predicted future utility, if they must be inserted in the SLLC or bypassed to the main memory. The Reuse Detector succeeds in managing the STT-RAM SLLC contents in two complementary ways. First, it is able to bypass to main memory a significant fraction of the blocks coming to the SLLC, thus decreasing the amount of the energy-hungry writes to be performed. Second, it increases significantly the SLLC hit rate, which leads to moderate performance improvements. In addition, the energy consumption in the main memory is also reduced. This way, our approach is able to outperform other strategies also oriented to decrease the energy consumption in STT-RAM SLLCs, such as the DASCA scheme. Although DASCA exhibits slightly lower ability to cut the write operations to the SLLCs, this technique, which predicts if a block will not be reused again instead of predicting if a block is going to be reused as ours, achieves lower accuracy in the prediction, hence also significantly lower hit rates at this cache level and therefore much lower performance improvements. Overall RD reports on average energy reductions in the SLLC in the range of 37-30\%, additional energy savings in the main memory in the range of 6-8\% and performance improvements of 3\% (quad-core), 7\% (eight-core) and 14\% (16-core) compared to an STT-RAM SLLC baseline where no reuse detector is employed. More importantly, our approach outperforms DASCA, the state-of-the-art STT-RAM SLLC management, reporting --depending on the specific scenario and the kind of applications used-- SLLC energy savings in the range of 4-11\% higher than those of DASCA, delivering higher performance in the range of 1.5-14\%, and an additional improvement in DRAM energy consumption in the range of 2-9\% higher than DASCA.


\balance

\ack{This work has been supported in part by the Spanish government through the research contracts TIN2012-32180, TIN2015-65277-R, TIN2015-65316-P and by the HIPEAC-4 European Network of Excellence. It has been also supported by a grant scholarship from the University of Costa Rica and the Costa Rican Ministry of Science and Technology MICIT and CONICIT.}


\bibliographystyle{compj}
\bibliography{acmsmall-sample-bibfile}

\end{document}